\documentclass[12pt]{article}
\usepackage{jheppub}

\usepackage{amsmath,amssymb,amsfonts}
\usepackage{graphicx}
\usepackage{color}
\usepackage{tikz}
\usepackage{dsfont}

\usepackage{booktabs}
\usepackage{multirow}

\usepackage{subfigure}

\usepackage{comment}

\def\Tr{\mbox{Tr}}

\def\gcd{{\rm gcd}}

\makeatletter\def\l@subsubsection#1#2{}%
\makeatother

\def\be{\begin{eqnarray}}
\def\ee{\end{eqnarray}}

\def\makeatletter{\catcode`\@=11}
\makeatletter
\def\mathbox#1{\hbox{$\m@th#1$}}%
\def\math@ccstyles#1#2#3#4#5#6#7{{\leavevmode
      \setbox0\mathbox{#6#7}%
      \setbox2\mathbox{#4#5}%
      \dimen@ #3%
      \baselineskip\z@\lineskiplimit#1\lineskip\z@
      \vbox{\ialign{##\crcr
             \hfil \kern #2\box2 \hfil\crcr
             \noalign{\kern\dimen@}%
             \hfil\box0\hfil\crcr}}}}
\def\mathaccstyles{\math@ccstyles\maxdimen}
\def\maththroughstyles{\math@ccstyles{-\maxdimen}}
\def\unity%
 {\maththroughstyles{.45\ht0}\z@\displaystyle {\mathchar"006C}\displaystyle 1}

\begin{document}

\title{The holography of duality in ${\cal N}=4$ Super-Yang-Mills theory}

\author[a]{Oren Bergman,}
\emailAdd{bergman@physics.technion.ac.il}
\author[b,c]{Shinji Hirano,}
\emailAdd{shinji.hirano@gmail.com}

\affiliation[a]{Department of Physics, Technion, Israel Institute of Technology\\
Haifa, 32000, Israel\\[-2mm]}

\affiliation[b]{School of Science, Huzhou University \\ Huzhou 313000, Zhejiang, China }  
\affiliation[c]{ Center for Gravitational Physics and Quantum Information
\\ Yukawa Institute for Theoretical Physics, Kyoto University
\\ Kitashirakawa-Oiwakecho, Sakyo-ku, Kyoto 606-8502, Japan}

\preprint{YITP-22-09}

\abstract{The space of ${\cal N}=4$ supersymmetric Yang-Mills theories exhibits an intricate structure of 
global one-form symmetries and $SL(2,\mathbb{Z})$ duality orbits. In this paper we study this structure from the point of view of the
holographic dual Type IIB string theory. Generalizing work by Witten, we map the different theories
based on the gauge algebras $su(N)$, $so(N)$, and $sp(N)$ to a choice of boundary conditions on bulk gauge fields.
We show how the one-form symmetries and their anomalies, as well as the duality properties of the gauge theories, arise in 
the holographic picture. Along the way we prove that the number of disjoint $SL(2,\mathbb{Z})$ duality orbits for 
the $su(N)$ theories is given by the number of square divisors of $N$.}

\maketitle

\section{Introduction}


One of the most interesting properties of ${\cal N}=4$ SYM theory is duality, namely the property that one ${\cal N}=4$ SYM theory with 
a given value of the complexified coupling parameter $\tau$ is physically equivalent to another, generically different, ${\cal N}=4$ SYM theory
with a different value of $\tau$, given by an action by an element of a discrete duality group on the original value.
For simply laced gauge groups the duality group is $SL(2,\mathbb{Z})$, which is generated by the GNO-duality $S$ transformation,
and the perturbative $T$ transformation that shifts the Yang-Mills theta angle by $2\pi$. For non-simply laced groups the duality group is slightly different.
The duality acts on the gauge algebra ${\bf g}$ in a simple way:
$T$ acts trivially, and $S$ maps the algebra ${\bf g}$ to the GNO-dual algebra ${\bf g}^*$.
On the other hand the action of the duality on the gauge theories turns out to be rather intricate \cite{Aharony:2013hda}.
For example the theory with $G=SU(N)$ is related by $S$ to the theory with $G=PSU(N)=SU(N)/\mathbb{Z}_N$.
More generally gauge theories based on the same algebra can differ in the global structure of the gauge group, as well as in values of discrete theta-like
parameters. The different theories have the same spectrum of local operators, but a different spectrum of extended line operators 
\cite{Aharony:2013hda},
and correspondingly a different global one-form symmetry \cite{Gaiotto:2014kfa}.
The different ${\cal N}=4$ gauge theories based on ${\bf g}$ and ${\bf g}^*$ are related by the duality, but generically
form several disjoint duality orbits \cite{Aharony:2013hda}.
In particular for ${\bf g}=su(N)$ with $N$ {\em square-free} there is only one $SL(2,\mathbb{Z})$ duality orbit,
but otherwise there are more.

The $AdS/CFT$ correspondence, the most famous example of which relates ${\cal N}=4$ SYM theory to Type IIB string theory in $AdS_5$,
provides an alternative and complementary point of view on these topics.
Soon after Maldacena's discovery of $AdS/CFT$  \cite{Maldacena:1997re},
Witten showed that different theories based on ${\bf g}=su(N)$ correspond to different
choices of boundary conditions for the NSNS and RR 2-form gauge fields $B_2$ and $C_2$ in $AdS_5$ \cite{Witten:1998wy}.
In particular any theory related by an $SL(2,\mathbb{Z})$ duality transformation to the theory with $G=SU(N)$ corresponds to a boundary condition on $(B_2,C_2)$
that is related by the same element of the $SL(2,\mathbb{Z})$ symmetry of Type IIB string theory to the boundary condition on $(B_2,C_2)$ corresponding to the $SU(N)$ theory.

In this paper we will extend Witten's results to include all the ${\cal N}=4$ SYM theories based on the $su(N)$, $so(2N)$, $so(2N+1)$,
and $sp(N)$ algebras. We will match all the different ${\cal N}=4$ gauge theories of \cite{Aharony:2013hda} to a corresponding set
of boundary conditions on the appropriate gauge fields in the dual $AdS_5$ background. Using the dual Type IIB string theory description 
we will reproduce the spectrum of line operators, the one-form symmetries and their anomalies, as well as the structure of the duality orbits.
Along the way we will also prove that the number of duality orbits for the theories based on $su(N)$ is given by the number
of square divisors of $N$.

The rest of the paper is organized according to the gauge algebras.
In section 2 we discuss the $su(N)$ theories. We begin by reviewing the results of \cite{Aharony:2013hda,Gaiotto:2014kfa} on the different theories and their 
one-form global symmetries, and how they are related by $SL(2,\mathbb{Z})$ duality. We also review two mixed anomalies in this class of theories 
following \cite{Hsin:2020nts} and \cite{Cordova:2019uob}. We then discuss the holographic description of the different theories in terms of the boundary conditions
imposed on the NSNS and RR 2-form fields in the bulk, and how the spectrum of line operators and one-form symmetries of the gauge theories 
are realized in terms of branes in the bulk. We also explain how the 5d anomaly actions corresponding to the mixed anomalies arise as Chern-Simons terms in the 
supergravity action.
Finally, we provide a brane realization, and the corresponding fully backreacted supergravity solution, of 
an interface in the $SU(N)$ gauge theory, separating regions with theta parameters differing by a multiple of $2\pi$.
In section 3 we discuss the $so(2n)$ theories, starting again with a review of the results of \cite{Aharony:2013hda} on the spectrum of line operators and 
the action of duality. We then move to the holographic description, find the dual boundary conditions, and discuss the brane realization of the line operators 
and one-form symmetries. In particular we highlight the distinction between $so(4k)$ and $so(4k+2)$, and the distinction between $so(8k)$ and $so(8k+4)$.
Section 4 contains the analogous discussion of the $so(2n+1)$ and $sp(n)$ theories.
Section 5 contains our conclusions and future directions.

\section{The $su(N)$ theories}
\label{su(N)}

\subsection{One form symmetry and duality}

YM theories with gauge algebra $su(N)$ are characterized in general by the global structure of the gauge group,
$SU(N)/\mathbb{Z}_k$, where $k$ is a divisor of $N$, and by the value of a discrete theta-like parameter 
$\ell$ taking values in  $\{0,\ldots,k-1\}$ \cite{Aharony:2013hda}.
The local properties of all the theories are the same, but the spectrum of line operators depends on $k$ and $\ell$:
\be
\label{LineOperatorSpectrum}
L_{k,\ell} = \{(z_e,z_m) = e(k,0) + m(\ell,k') \; \mbox{mod}\, N\} \,,
\ee
where $k' \equiv N/k$ and $e,m\in \mathbb{Z}$.
A priori the charges $z_e,z_m$ take values in the $\mathbb{Z}_N$ center of $su(N)$, but they are constrained by 
mutual locality via the Dirac pairing condition
\be
\label{DiracPairing}
z_{e}z_{m}' - z_{m}z_e' = 0 \; \mbox{mod} \; N \,.
\ee
It was argued in \cite{Aharony:2013hda} that for a consistent quantum field theory the set of charges must be maximal and complete.
The different maximal sets can be represented as $L_{k,\ell}$.

The different theories are related by $SL(2,\mathbb{Z})$ duality in a non-trivial way \cite{Aharony:2013hda}.
The $T$ generator of $SL(2,\mathbb{Z})$ takes $(z_e,z_m) \rightarrow (z_e,z_m+z_e)$, and therefore
\be
\label{GaugeTtrans}
(k,\ell) \stackrel{T}{\longrightarrow} (k,\ell + k') \,.
\ee
The $S$ generator takes $(z_e,z_m) \rightarrow (z_m,-z_e)$, which implies that
\be
\label{GaugeStrans}
(k,\ell) \stackrel{S}{\longrightarrow} \left(\frac{N}{\mbox{gcd}(k,\ell)},m_0k'\right) \,,
\ee
where $m_0$ is a member of a Bezout pair of integers $(e_0,m_0)$ 
associated to the pair $(k,\ell)$.\footnote{This is determined by finding the minimal charges in $L_{k,\ell}$ of the form
$(0,k^*)$ and $(-{k'}^*,\ell^*)$. For $(0,k^*)$ these are given by $(e_0,m_0) = (-\ell,k)/\mbox{gcd}(k,\ell)$,
and therefore $k^* = N/\mbox{gcd}(k,\ell)$. Then for $(-{k'}^*,\ell^*)$ we have $m_0 k' = \ell^*$ and $e_0 k + m_0 \ell = -{k'}^* = -\mbox{gcd}(k,\ell)$,
which implies that $(e_0,m_0)$ are a Bezout pair associated to $(k,\ell)$.}
For example $S$ maps the $SU(N)$ theory to the $(SU(N)/\mathbb{Z}_N)_0$ theory,
and then acting with $T$ leads to all the other $(SU(N)/\mathbb{Z}_N)_\ell$ theories.
However in general not all theories are related by $SL(2,\mathbb{Z})$ to the $SU(N)$ theory.
This is only the case if $N$ is {\em square-free}, {\em i.e.} if all of its prime factors appear only once \cite{Aharony:2013hda}.
More generally the theories form different orbits under $SL(2,\mathbb{Z})$. 
As we will show below, the number of different orbits is equal to the number of different square divisors of $N$,
and is therefore given by
\be 
n_{orbits} = \prod_{i} \left(1+ \left[\frac{e_i}{2}\right]\right) \,,
\ee
where $[e_i/2]$ is the integer part of $e_i/2$, and $e_i$ is the exponent of the prime $p_i$ in the prime factorization of $N$,
\be
N=\prod_{i} p_i^{e_i} \,.
\ee
For a square-free integer $N$ all $e_i$ are either 0 or 1, and therefore $n_{orbits}=1$.

The structure of the different theories can also be described in terms of higher form symmetry \cite{Gaiotto:2014kfa}.
The $SU(N)$ theory has a $\mathbb{Z}^{(1)}_N$ global one-form symmetry, corresponding to the center of $SU(N)$, 
acting on ``electric" Wilson lines in arbitrary representations of $SU(N)$.
The $\mathbb{Z}_N^{(1)}$ charge of a Wilson line is given by the $N$-ality of the representation, 
{\em i.e.} the number of boxes in the corresponding Young diagram.
The $SU(N)/\mathbb{Z}_k$ theory is obtained by gauging a $\mathbb{Z}^{(1)}_k$ subgroup of $\mathbb{Z}^{(1)}_N$.
This removes electric lines with charges that are not a multiple of $k$, leaving an electric one-form symmetry $\mathbb{Z}^{(1)}_{k'}$
acting on the remaining electric lines.
The new theory also admits additional line operators carrying magnetic charges that are multiples of $k'$.
If these lines carry no electric charge then they are acted on by a magnetic one-form symmetry $\mathbb{Z}^{(1)}_k$,
and the total one-form symmetry is $\mathbb{Z}^{(1)}_{k'}\times \mathbb{Z}^{(1)}_{k}$.
If $k$ and $k'$ are mutually prime this is equivalent to $\mathbb{Z}_N^{(1)}$.
More generally the spectrum of line operators is given by (\ref{LineOperatorSpectrum}), 
and the one-form symmetry of the $(SU(N)/\mathbb{Z}_k)_\ell$ theory is given by
\be
\label{OneFormSymmetry1}
G^{(1)} = (\mathbb{Z}_{k'} \times \mathbb{Z}_{N/gcd(k',\ell)})/\mathbb{Z}_{k'/gcd(k',\ell)} \,.
\ee
The first factor $\mathbb{Z}_{k'}$ 
acts on the electric lines $e(k,0)$, the second factor $\mathbb{Z}_{N/gcd(k',\ell)}$
acts on the dyonic lines $m(\ell,k')$, 
and the quotient accounts for the relation $\ell(k,0) = k(\ell,k')$ mod $N$.
This can be more simply expressed as 
\be
\label{OneFormSymmetry2}
G^{(1)} =   \mathbb{Z}_{N/gcd(k,k',\ell)}  \times  \mathbb{Z}_{gcd(k,k',\ell)} \,, 
\ee
by bringing the matrix of fundamental line charges to Smith normal form:
\be
\label{SmithNormalForm}
U\left(
\begin{array}{cc}
k & 0 \\
\ell & k'
\end{array}\right)V^{-1} = \left(
\begin{array}{cc}
\mbox{gcd}(k,k',\ell) & 0 \\
0 & N/\mbox{gcd}(k,k',\ell)
\end{array} \right) \,,
\ee
where $U,V$ are $SL(2,\mathbb{Z})$ matrices.\footnote{The explicit transformation is presented in Appendix A.}
For $\ell=0$ this reduces to 
$\mathbb{Z}_{N/gcd(k,k')}  \times  \mathbb{Z}_{gcd(k,k')} = \mathbb{Z}_{k'} \times \mathbb{Z}_k$.

The transformation in (\ref{SmithNormalForm}) shows that any theory of the form $(SU(N)/\mathbb{Z}_k)_\ell$ 
may be mapped by $SL(2,\mathbb{Z})$ duality to the theory $(SU(N)/\mathbb{Z}_d)_0$, 
with $d=\mbox{gcd}(k,k',\ell)$.\footnote{Indeed the quantity $\mbox{gcd}(k,k',\ell)$ is invariant under $SL(2,\mathbb{Z})$.
Under the $T$ transformation 
$\mbox{gcd}(k,k',\ell) \stackrel{T}{\longrightarrow} \mbox{gcd}(k,k',\ell +k') = \mbox{gcd}(k,k',\ell)$,
and under the $S$ transformation
$\mbox{gcd}(k,k',\ell)  \stackrel{S}{\longrightarrow}  
\mbox{gcd}(N/\mbox{gcd}(k,\ell),\mbox{gcd}(k,\ell),mk') 
= \mbox{gcd}(k'\mbox{gcd}(k/\mbox{gcd}(k,\ell),m),\mbox{gcd}(k,\ell)) 
= \mbox{gcd}(k',\mbox{gcd}(k,\ell)) 
= \mbox{gcd}(k,k',\ell)$,
where we have used the fact that $nk - m\ell = \mbox{gcd}(k,\ell)$ implies that that the integers $m$ and $k/\mbox{gcd}(k,\ell)$ 
are a co-prime pair.}
The different $SL(2,\mathbb{Z})$ orbits are therefore classified by the possible values of $d$.
For example for $N=4$ there are two separate $SL(2,\mathbb{Z})$ orbits of theories corresponding to
$d = 1$ and $d = 2$.
The former contains the six theories $SU(4)$, $(SU(4)/\mathbb{Z}_4)_{0,1,2,3}$, and $(SU(4)/\mathbb{Z}_2)_1$,
and the latter contains just the theory $(SU(4)/\mathbb{Z}_2)_0$  \cite{Aharony:2013hda}.
The number of orbits for $su(N)$ is given by the number of distinct values that the integer $d$ can have.
Since we can take the seed theory for each orbit to be $(SU(N)/\mathbb{Z}_d)_0$, this translates to the number 
of different divisors $d$ of $N$ satisfying $d = \mbox{gcd}(d,N/d)$, or equivalently to the number of square divisors $d^2$ of $N$.

\subsection{Anomalies}

There is a mixed anomaly between the $\mathbb{Z}_d$ and $\mathbb{Z}_{N/d}$  factors
in the one-form symmetry (\ref{OneFormSymmetry2}) determined by the extension
\be
1 \longrightarrow \mathbb{Z}_{d} \longrightarrow \mathbb{Z}_N \longrightarrow \mathbb{Z}_{N/d} \longrightarrow 1 \,.
\ee
The degree of the anomaly is equal to the degree of the extension, which is $\mbox{gcd}(d,N/d) = d = \mbox{gcd}(k,k',\ell)$.
For the special case of $\ell = 0$ this is a mixed anomaly between the electric $\mathbb{Z}_{k'}$ one-form symmetry and the
magnetic $\mathbb{Z}_k$ one-form symmetry \cite{Gaiotto:2014kfa}. The 5d anomaly action in this case is given by \cite{Hsin:2020nts} 
\be
\label{AnomalyAction1}
S_{anomaly_1} = \frac{2\pi}{N} \int_Y \mathsf{C} \cup \delta \mathsf{B}
\ee
where $\mathsf{B}$ and $\mathsf{C}$ are background fields of the electric and magnetic symmetries, 
respectively, namely
\be 
\mathsf{B} \in H^2(X,\mathbb{Z}_{k'}) \; , \; \mathsf{C} \in H^2(X,\mathbb{Z}_{k}) \,.
\ee
The order of the anomaly in this case is $\mbox{gcd}(k,k')$.
We can see this explicitly using the fact that we can write $\mbox{gcd}(k,k')=nk + mk'$ for some pair of integers $(n,m)$,
and multiplying the anomaly action by this quantity. Integrating the second term by parts gives
\be
\mbox{gcd}(k,k') \cdot S_{anomaly_1} =  \frac{2\pi n}{k'} \int_Y \mathsf{C} \cup \delta \mathsf{B} -  \frac{2\pi m}{k} \int_Y \mathsf{B} \cup \delta \mathsf{C} \,,
\ee
and since $\delta \mathsf{B} = 0 \; \mbox{mod} \; k'$ and  $\delta \mathsf{C} = 0 \; \mbox{mod} \; k$, this is a multiple of $2\pi$.

More generally we have $\delta \mathsf{B} = 0 \; \mbox{mod} \; k'$  
and $k'\delta\mathsf{C} + \ell\delta\mathsf{B} = 0 \; \mbox{mod} \; N$.
If $\ell \neq 0$  the order of the anomaly is reduced to $\mbox{gcd}(k,k',\ell)$. This can be seen by performing the following change of basis
\be 
\mathsf{C}' &=& \frac{pk + \ell}{\gcd(k,k',\ell)} \mathsf{B} + \frac{k'}{\gcd(k,k',\ell)} \mathsf{C} \\
\mathsf{B}' &=& \beta \mathsf{B} + \alpha \mathsf{C}
\ee
where $p$ is an integer satisfying
\be
\mbox{gcd}(k',pk+\ell) = \mbox{gcd}(k,k',\ell) \,,
\ee
and $(\alpha,\beta)$ is a pair of integers satisfying
\be
{k'\beta\over\gcd(k,k',\ell)}-{(pk+\ell)\alpha\over\gcd(k,k',\ell)}=1\ .
\ee
(See the Appendix A for the proof of their existence).
It follows that
\be
\delta\mathsf{C}' & = &0\,\,\,\,\mbox{mod}\,\,\,\, N/\gcd(k,k',\ell)  \ ,\\
\delta\mathsf{B}' & = & 0\,\,\,\,\mbox{mod}\,\,\,\, \gcd(k,k',\ell)  \, .
\ee
We can then express the anomaly action as
\begin{align}
S_{\rm anomaly_1}={2\pi\over N}\int_Y \mathsf{C}'\cup \delta\mathsf{B}'
-\underbrace{{2\pi\over 2N}\int_X\left[{(pk+\ell)\beta\over \gcd(k,k',\ell)}\mathsf{B}\cup\mathsf{B}
+{k'\alpha\over \gcd(k,k',\ell)}\mathsf{C}\cup\mathsf{C}\right]}_{\rm canceled\,\,by\,\,counterterms} \,,
\end{align}
where the last two terms can be cancelled by 4d counterterms.
This has the same form as (\ref{AnomalyAction1}) with $\mathsf{B},\mathsf{C}$ replaced by $\mathsf{B}',\mathsf{C}'$.
Therefore the order of the anomaly is $\gcd(\gcd(k,k',\ell), N/\gcd(k,k',\ell))=\gcd(k,k',\ell)$. 

There is another kind of mixed anomaly in these theories that involves the electric one-form symmetry and 
the periodicity of the $\theta$ parameter  \cite{Cordova:2019uob}.
For example in the $SU(N)$ theory the $\theta$ parameter has a $2\pi$ periodicity, $\theta \sim \theta + 2\pi$.
 Turning on a background field $\mathsf{B}$ for the electric one-form symmetry $\mathbb{Z}_N$
 leads to a phase in the path integral under $\theta \rightarrow \theta + 2\pi$,
 \be 
Z[\theta + 2\pi] = Z[\theta] \exp\left[{2\pi i \, \frac{N-1}{N}\int_X \frac{{\cal P}(\mathsf{B})}{2}}\right] \,.
\ee
A non-trivial $\mathsf{B}$ corresponds to a fractional instanton, an $SU(N)/\mathbb{Z}_N$ bundle that
cannot be lifted to an $SU(N)$ bundle.
Gauging the one-form symmetry, by summing over all elements $\mathsf{B} \in H^2(X,\mathbb{Z}_N)$,
would then violate the $2\pi$ periodicity of $\theta$.\footnote{For $\theta=\pi$ this also implies that
there is a mixed anomaly between the one-form symmetry and time reversal \cite{Gaiotto:2017yup}.}
We can express this anomaly in terms of a 5d anomaly action given by
\be
\label{AnomalyAction2}
S_{anomaly_2} = 2\pi i \, \frac{N-1}{N}\int_Y \frac{d\theta}{2\pi} \cup \frac{{\cal P}(\mathsf{B})}{2} \,.
\ee
Note that the ${\cal O}(1)$ part of the anomaly, namely the $N$ in the numerator, is trivial for spin manifolds,
since $\int {\cal P}(\mathsf{B})$ is an even integer for spin manifolds.
The same anomaly action (\ref{AnomalyAction2}) holds more generally for the $(SU(N)/\mathbb{Z}_k)_\ell$ theory.
In this case the periodicity of the $\theta$ parameter increases to $2\pi k$, and the analogous phase in the path integral
is given by 
\be 
Z[\theta + 2\pi k] = Z[\theta] \exp\left[{2\pi i \, \frac{N-1}{k'}\int_X \frac{{\cal P}(\mathsf{B})}{2}}\right] \,,
\ee
where now $\mathsf{B} \in H^2(X,\mathbb{Z}_{k'})$. Note that $\mathsf{B}$ is still the background field of the electric
one-form symmetry, which is the $\mathbb{Z}_{k'}$ subgroup of (\ref{OneFormSymmetry1}) or (\ref{OneFormSymmetry2}).

\subsection{Holography}

The holographic dual of ${\cal N}=4$ SYM theory with gauge algebra $su(N)$ is Type IIB string theory on $AdS_5\times S^5$ \cite{Maldacena:1997re}.
The global structure of the theory depends on the boundary conditions satisfied by the NSNS and RR 2-forms $B_2, C_2$ 
\cite{Witten:1998wy}.
These boundary conditions are constrained by a 5d Chern-Simons term 
\be
\label{IIBAction}
S_{CS}[B_2,C_2] = \int_{AdS_5\times S^5} B_2 \wedge dC_2 \wedge dC_4 = \frac{N}{2\pi} \int_{AdS_5} B_2\wedge dC_2 \,,
\ee
which is the dominant term near the boundary of $AdS_5$.
This implies that the boundary values of $B_2$ and $C_2$ are flat, and therefore completely characterized by their holonomies
on closed two-surfaces at the boundary,\footnote{What one should really have in mind is a topologically non-trivial boundary with homology 2-cycles, such 
as $S^1\times S^3/\mathbb{Z}_k$.}
\be
e^{ib(S)} & = & e^{i \oint_S B_2} \\
e^{ic(S)} & = & e^{i\oint_S C_2} \,.
\ee
The CS action also implies that $B_2$ and $C_2$ are quantum-mechanically conjugate variables, and in particular that
\be
e^{ic(S)} e^{ib(S')} = e^{ib(S')} e^{ic(S)} \exp\left[\left(\frac{2\pi i}{N}\right) S \cdot S'\right] \,,
\ee
where $S\cdot S'$ is the intersection number of the surfaces $S$ and $S'$.
In other words $b$ generates a $\mathbb{Z}_N$ ``translation" symmetry acting on $c$ as $c\rightarrow c + 2\pi/N$, and vice versa.
The quantum mechanical variables $b$ and $c$ can therefore be viewed as discrete position-like and momentum-like variables taking values in $\mathbb{Z}_N$:
\be
b,c \in \left\{0,\frac{2\pi}{N},\frac{4\pi}{N},\ldots,\frac{2(N-1)\pi}{N}\right\} \,,
\ee
with 
\be
[b,c] = \frac{2\pi i}{N} \;\; \mbox{mod} \;\; 2\pi i\,.
\ee

The different boundary theories correspond to a choice of a maximal set of commuting observables.
The set must be maximal, since every quantum state is a simultaneous eigenstate of a maximal set
of commuting observables.\footnote{This can be shown as follows. Let $|A\rangle$ be a simultaneous
eigenstate of a maximal set of commuting observables ${\cal M}_i$, namely ${\cal M}_i |A\rangle = m_i |A\rangle$.
Let $|B\rangle$ be an arbitrary state in the Hilbert space. Then there exists a unitary operator ${\cal U}$ such that
$|B\rangle = {\cal U} |A\rangle$. We then have
${\cal U} {\cal M}_i {\cal U}^\dagger |B\rangle = {\cal U} {\cal M}_i |A\rangle = m_i {\cal U}|A\rangle = m_i|B\rangle$,
and therefore the state $|B\rangle$ is a simultaneous eigenstate of the maximal set of commuting observables
${\cal U} {\cal M}_i {\cal U}^\dagger$. We thank Shlomo Razamat for the proof.}
For example if we choose to fix $b$ at the boundary we can only simultaneously fix $Nc$,
which means that $c$ is free as an element of $\mathbb{Z}_N$.
The state in this case is dual to the $SU(N)$ theory.
More generally the set of observables of the form 
\be
{\cal O}_{n_b,n_c}(b,c) = n_b b + n_c c \,,
\ee
that we can simultaneously fix at the boundary must satisfy
\be
n_b n_c' - n_c n_b' = 0 \; \mbox{mod} \; N \,.
\ee
This is precisely the Dirac pairing condition of (\ref{DiracPairing}).
In other words the condition of mutual locality of the line operators in the boundary theory
corresponds to the condition of mutual commutativity of the observables in the bulk.
Furthermore, the requirement that the the spectrum of line operators be maximal follows from the requirement
that the set of observables be maximal.
The maximal set has at most two independent observables satisfying $n_b n_c' - n_c n_b' = N$.
By a convenient choice of basis we can set
\be
\label{SUBC}
(n_b,n_c) = (k,0) \; , \; (n'_b,n'_c) = (\ell,k') \,,
\ee
where $N=kk'$ and $0\leq \ell \leq k-1$. 
This pair of observables  corresponds to the $(SU(N)/\mathbb{Z}_k)_\ell$ theory.
In particular we can read off the one form symmetry (\ref{OneFormSymmetry1}) from the observables that are fixed at the boundary.
The observable ${\cal O}_{k,0}$ gives the $\mathbb{Z}_{k'}$ factor, the observable ${\cal O}_{\ell,k'}$ 
gives the $\mathbb{Z}_{N/gcd(k',\ell)}$ factor, and the relation $\ell{\cal O}_{k,0} = k{\cal O}_{\ell,k'}$ gives the quotient
by $\mathbb{Z}_{k'/gcd(k',\ell)}$.

Once we have identified the boundary conditions dual to the gauge theory, the action of $SL(2,\mathbb{Z})$ on the gauge theory
follows from the action of $SL(2,\mathbb{Z})$ in Type IIB string theory.
The $T$ generator takes $(b,c) \rightarrow (b,c+b)$ and therefore
\be
{\cal O}_{k,0}  \stackrel{T}{\longrightarrow} {\cal O}_{k,0} \; , \; {\cal O}_{\ell,k'}  \stackrel{T}{\longrightarrow} {\cal O}_{\ell+k',k'} \,,
\ee
which reproduces (\ref{GaugeTtrans}).
The $S$ generator takes $(b,c) \rightarrow (c,-b)$ and therefore
\be
{\cal O}_{k,0}  \stackrel{S}{\longrightarrow} {\cal O}_{0,k} \; , \; {\cal O}_{\ell,k'}  \stackrel{S}{\longrightarrow} {\cal O}_{-k',\ell} \,.
\ee
We can recast these in the form of (\ref{GaugeStrans}) by a change of basis
\be
\label{BasisChange2}
\left(
\begin{array}{cc}
\frac{\ell}{gcd(k,\ell)} &  \frac{-k}{gcd(k,\ell)}  \\
e_0 & -m_0
\end{array}\right)
\left(
\begin{array}{cc}
0 & k \\
- k' & \ell
\end{array}\right) = \left(
\begin{array}{cc}
\frac{N}{gcd(k,\ell)} & 0 \\
m_0 k' & \mbox{gcd}(k,\ell)
\end{array} \right) \,,
\ee
where $(e_0,m_0)$ are a Bezout pair associated to $(k,\ell)$.

\subsection{Holographic anomaly actions}

Anomaly actions are a useful way of encoding anomalies in terms of ``anomaly inflow" from a fictional space with one more dimension.
In holography this space is real, and the anomaly actions are part of the bulk physics.
They are given by Chern-Simons terms in the bulk.
In our case the relevant CS term is (\ref{IIBAction}).
The Type IIB 2-form gauge fields $B_2$ and $C_2$ are the continuum versions of the background fields $\mathsf{B}$ and 
$\mathsf{C}$. Identifying $B_2 =  2\pi \mathsf{B}/N$ and $C_2=2\pi \mathsf{C}/N$, we see that the 5d CS action (\ref{IIBAction})
reproduces the 5d anomaly action of the first mixed anomaly (\ref{AnomalyAction1}).

The second mixed anomaly involves the $\theta$ parameter, which is realized in the bulk in terms of the RR scalar field $C_0$.
There is no bulk CS term in Type IIB supergravity that involves this field directly.
However $C_0$ couples indirectly to the NSNS 2-form $B_2$ via the modified RR 3-form field strength,
\be
\tilde{F}_3 = dC_2 - C_0 dB_2 \,.
\ee
It is this combination that appears in the kinetic term. This allows us to rewrite the CS term in (\ref{IIBAction}) as
\be
S_{CS} 
= \mbox{} - \frac{N}{4\pi} \int_{AdS_5} dC_0 \wedge B_2\wedge B_2 + 
 \frac{N}{2\pi} \int_{AdS_5} \tilde{F}_3\wedge B_2 \, .
\ee
The first term reproduces the ${\cal O}(1/N)$ part of the anomaly action in (\ref{AnomalyAction2}).
The ${\cal O}(1)$ part is missing.
On the other hand, since we are considering a supersymmetric theory, and are therefore restricted to spin manifolds, 
the ${\cal O}(1)$ part of the anomaly is anyway trivial.

\subsection{Branes, lines, and surfaces}

Line operators in the 4d gauge theory correspond to one-dimensional boundaries of string worldsheets that end on the boundary of $AdS_5$.
The 4d theory also has two-dimensional surface operators described in the bulk by string worldsheets approaching, and parallel to, the boundary.
These surface operators implement the one-form symmetries acting on the line operators.
The spectrum of line and surface operators, and correspondingly the one-form symmetry group, depends crucially on the boundary conditions on $(b,c)$.

\begin{figure}[h!]
\center
\includegraphics[height=0.2\textwidth]{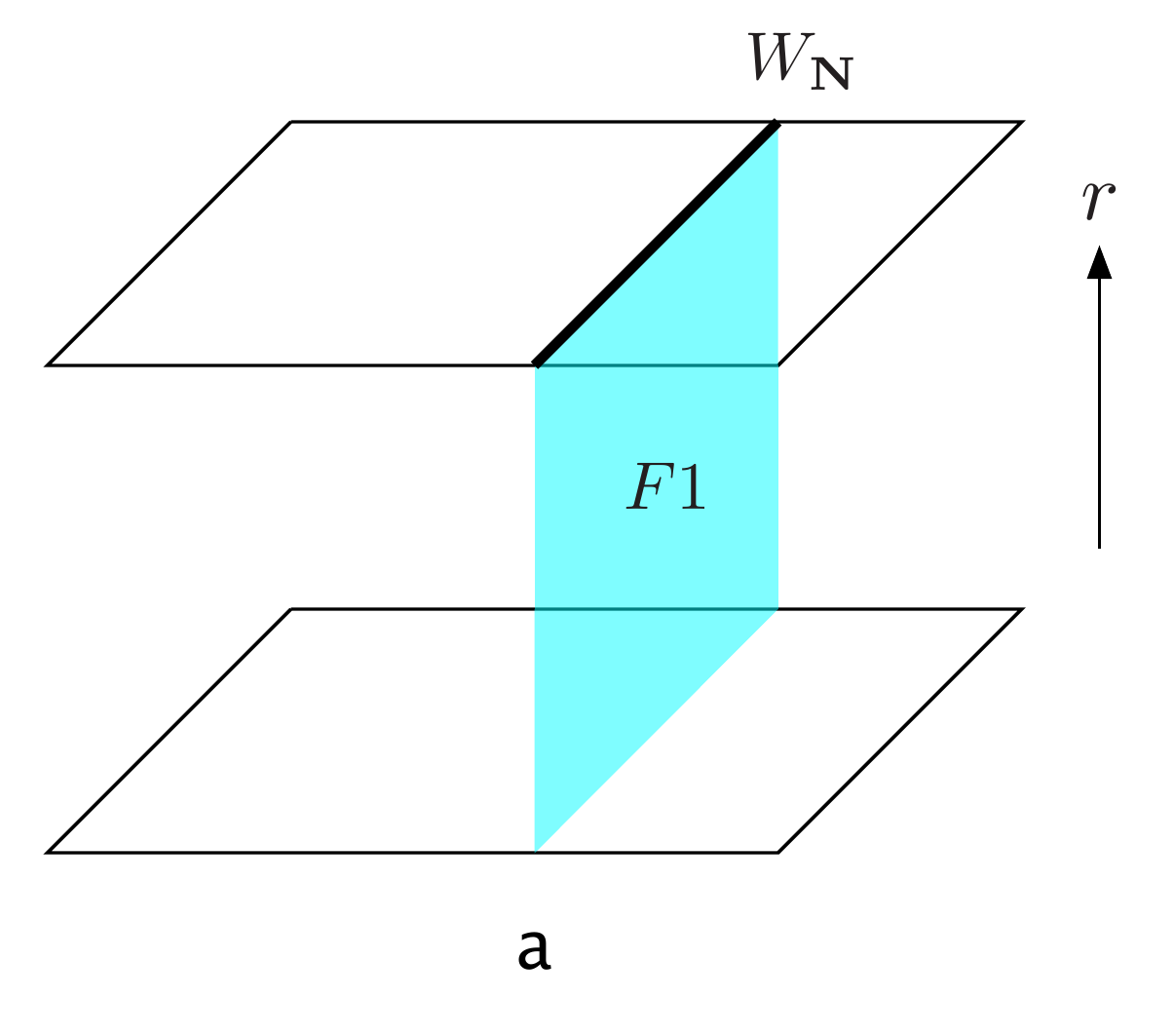}
\hspace{10pt} 
\includegraphics[height=0.2\textwidth]{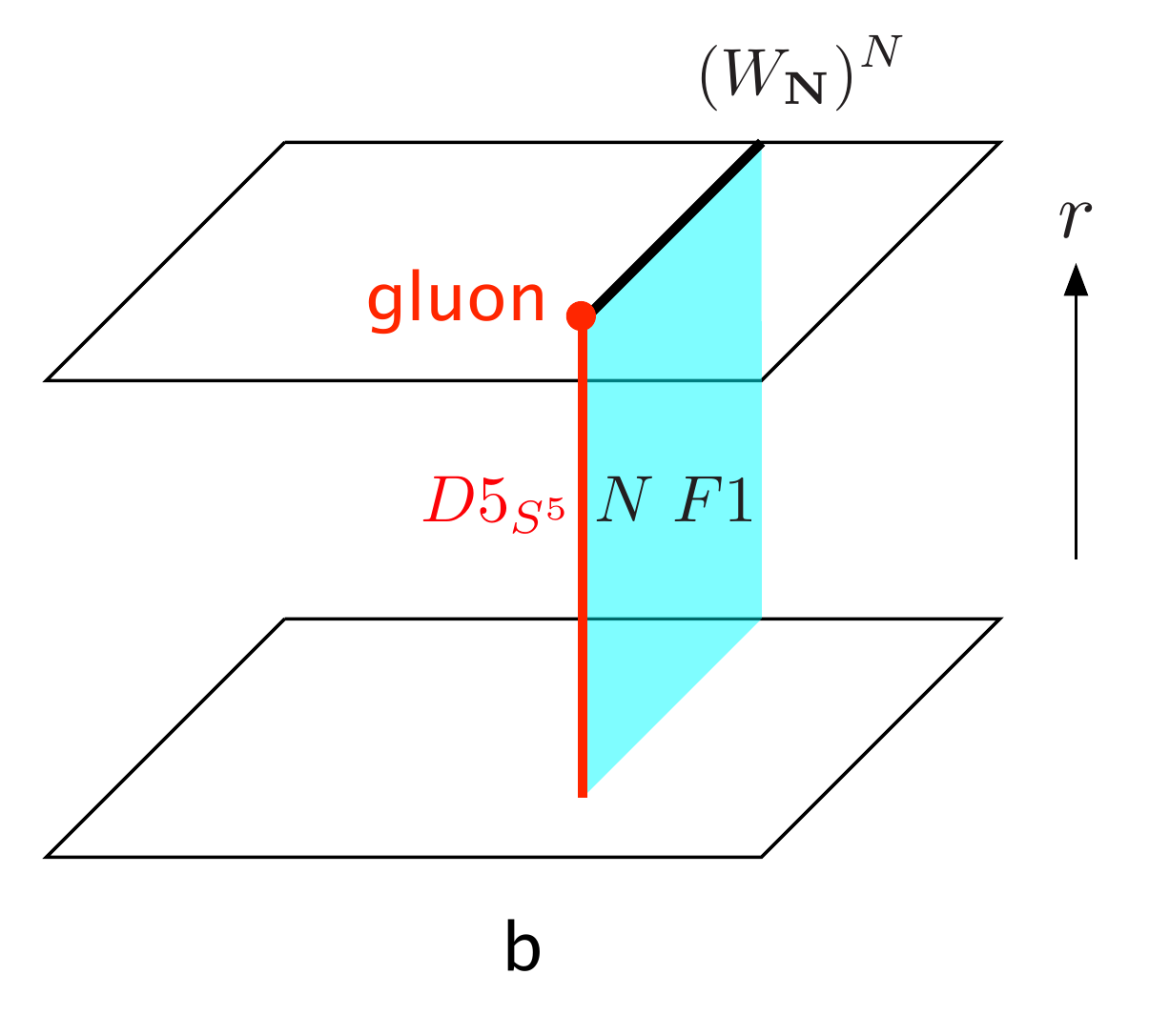} 
\hspace{10pt}
\includegraphics[height=0.2\textwidth]{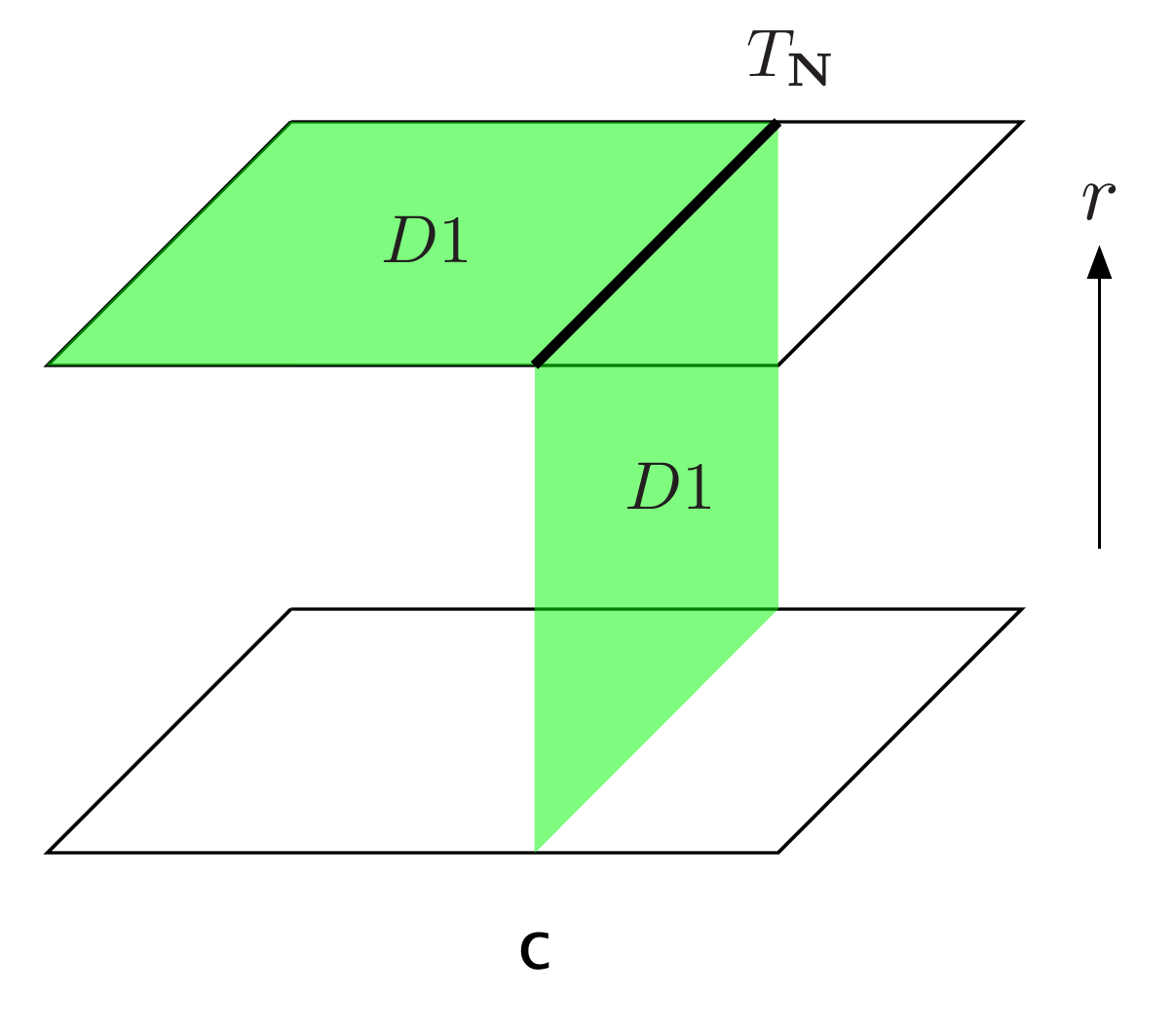} 
\caption{Bulk description of line operators in the $SU(N)$ theory: (a) A fundamental Wilson line, (b) $N$ fundamental Wilson lines screened by a gluon, (c) a fundamental 'tHooft line
attached to a Dirac-surface.}
\label{LinesSU(N)}
\end{figure}

Consider for example the $SU(N)$ theory.
The fixed boundary condition on the NSNS holonomy $b$ implies that the worldsheet of a fundamental string (or F-string) at the boundary corresponds to a trivial surface operator.
On the other hand the free boundary condition on the RR holonomy $c$ 
implies that the worldsheet of a D-string at the boundary describes a non-trivial surface operator 
acting on the line operators described by F-strings ending on the boundary, Fig.~\ref{LinesSU(N)}a.\footnote{A more precise description of a surface operator is given by an $AdS_3\times S^1$ D3-brane with magnetic flux perpendicular to the 2-surface which represents a D-string \cite{Gukov:2006jk, Drukker:2008wr}.}
That the one-form symmetry is $\mathbb{Z}_N$  is seen in the bulk by the screening of $N$ F-strings ending on the boundary by a Euclidean 
D5-brane wrapping $S^5$ and ending on the boundary, Fig.~\ref{LinesSU(N)}b.\footnote{One might be concerned with the possibility that this D5-brane might instead attach to a D5-brane at the boundary.
However the free boundary condition on $c$, which allows D-strings to approach the boundary, forbids wrapped D5-branes from doing so.
This is because Hodge duality exchanges free and fixed boundary conditions. Defining $A_1 = \int_{S^5} C_6$ as the 5d gauge field
that couples to the wrapped D5-brane, we have
\be
d_n A_1 = *_5 d_t C_2 \;\; , \;\; d_t A_1 = *_5 d_n C_2 \,,
\ee
where $d_n$ denotes the derivative normal to the boundary, and $d_t$ the derivative tangent to the boundary.}
This is a Euclidean version of Witten's Baryon vertex \cite{Witten:1998xy}.
In some sense it identifies the wrapped Euclidean D5-brane as the gluon operator of the 4d gauge theory.
D-strings ending on the boundary do not give rise to genuine line operators, since they can attach to a D-string surface at the boundary, Fig.~\ref{LinesSU(N)}c.
For the S-dual theory $SU(N)/\mathbb{Z}_N$ the situation is very similar, with the roles of the F-string and D-string exchanged.
Indeed we get basically the same thing for any $SL(2,\mathbb{Z})$ transform of the $SU(N)$ theory:
the non-trivial line operators are described by worldsheets of $(p,q)$-strings ending on the boundary, and $N$ such strings 
are screened by a wrapped Euclidean $(p,q)$ 5-brane ending on the boundary.

\begin{figure}[h!]
\center
\includegraphics[height=0.2\textwidth]{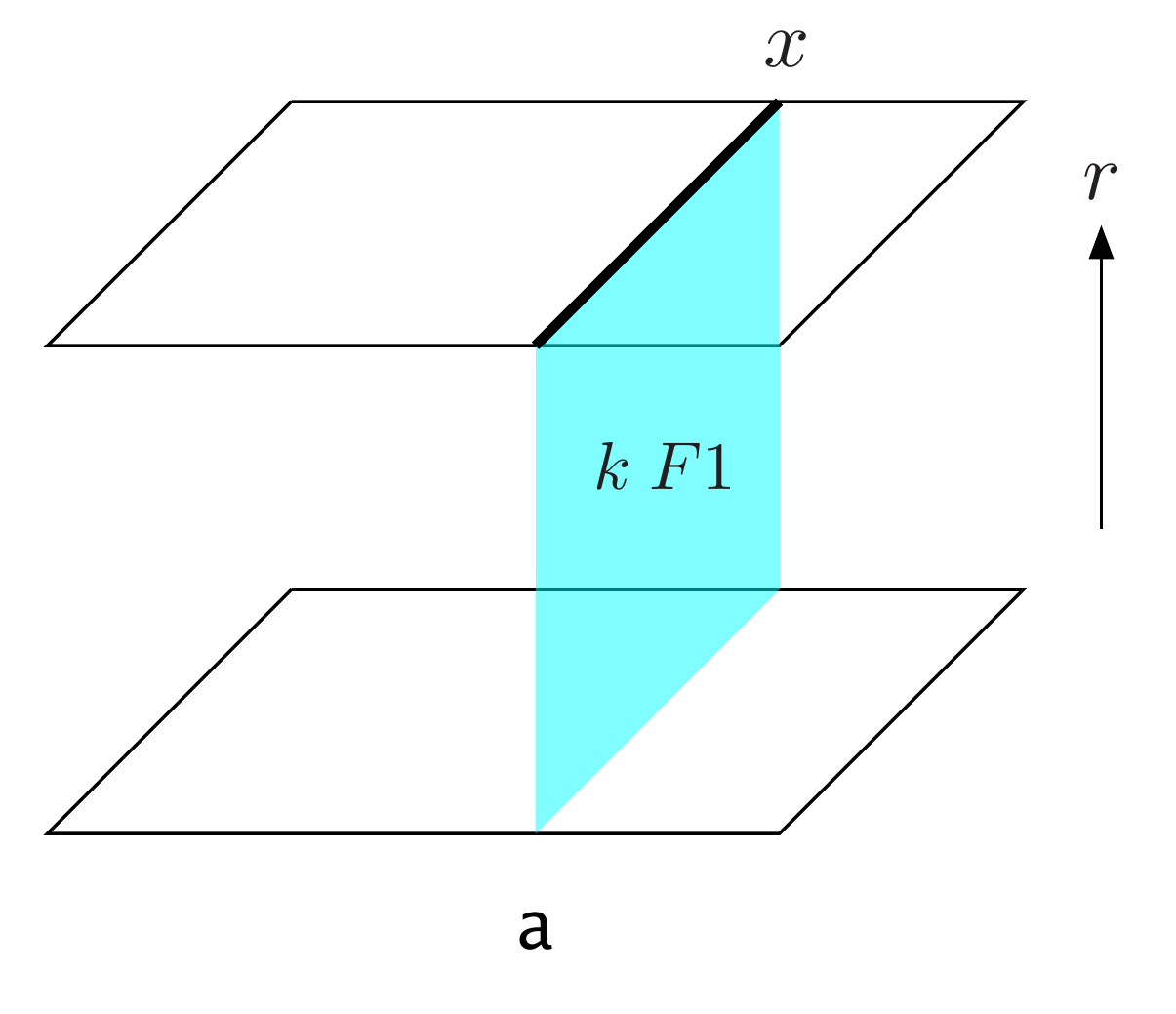} 
\hspace{10pt} 
\includegraphics[height=0.2\textwidth]{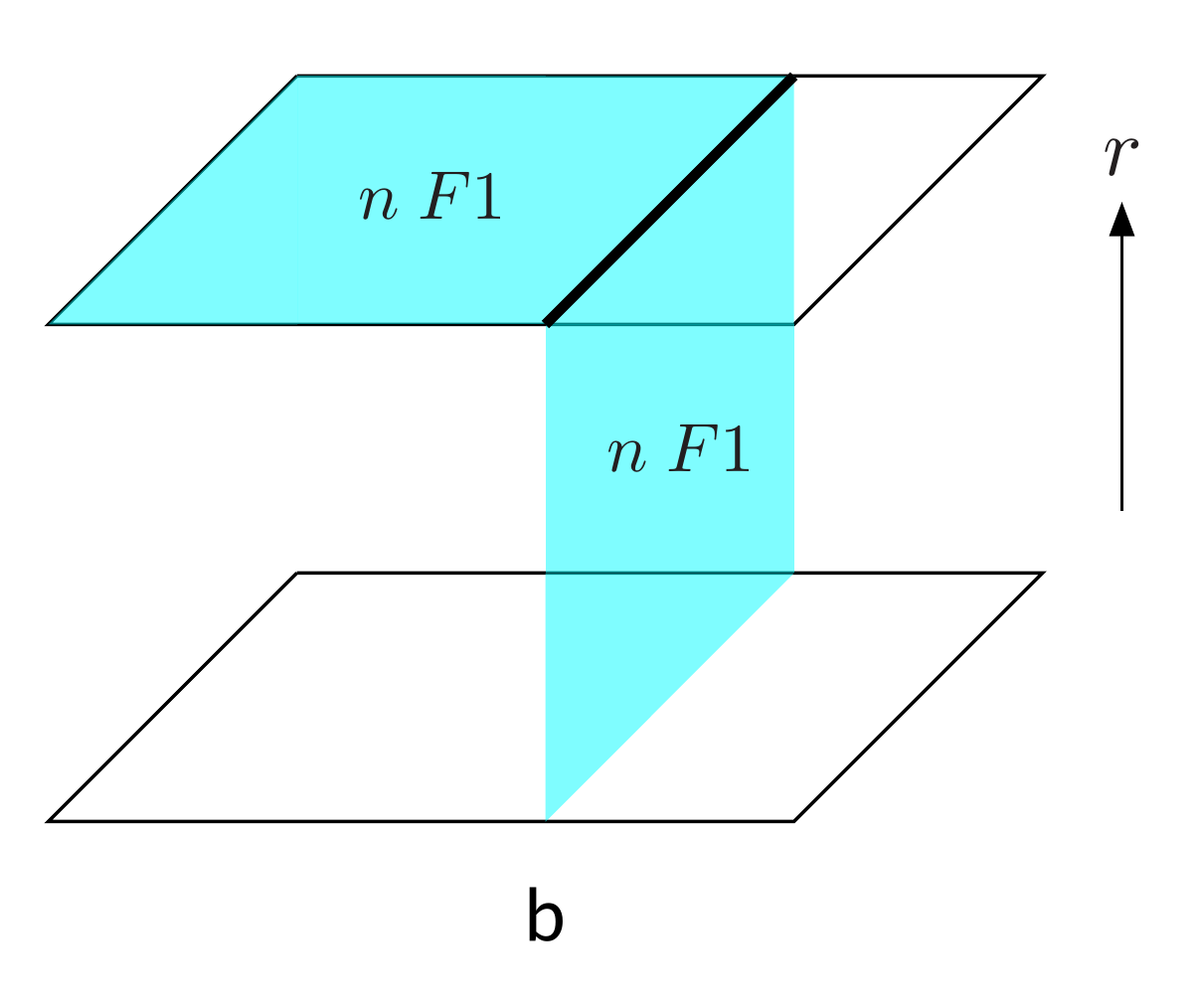}
\hspace{10pt}
\includegraphics[height=0.2\textwidth]{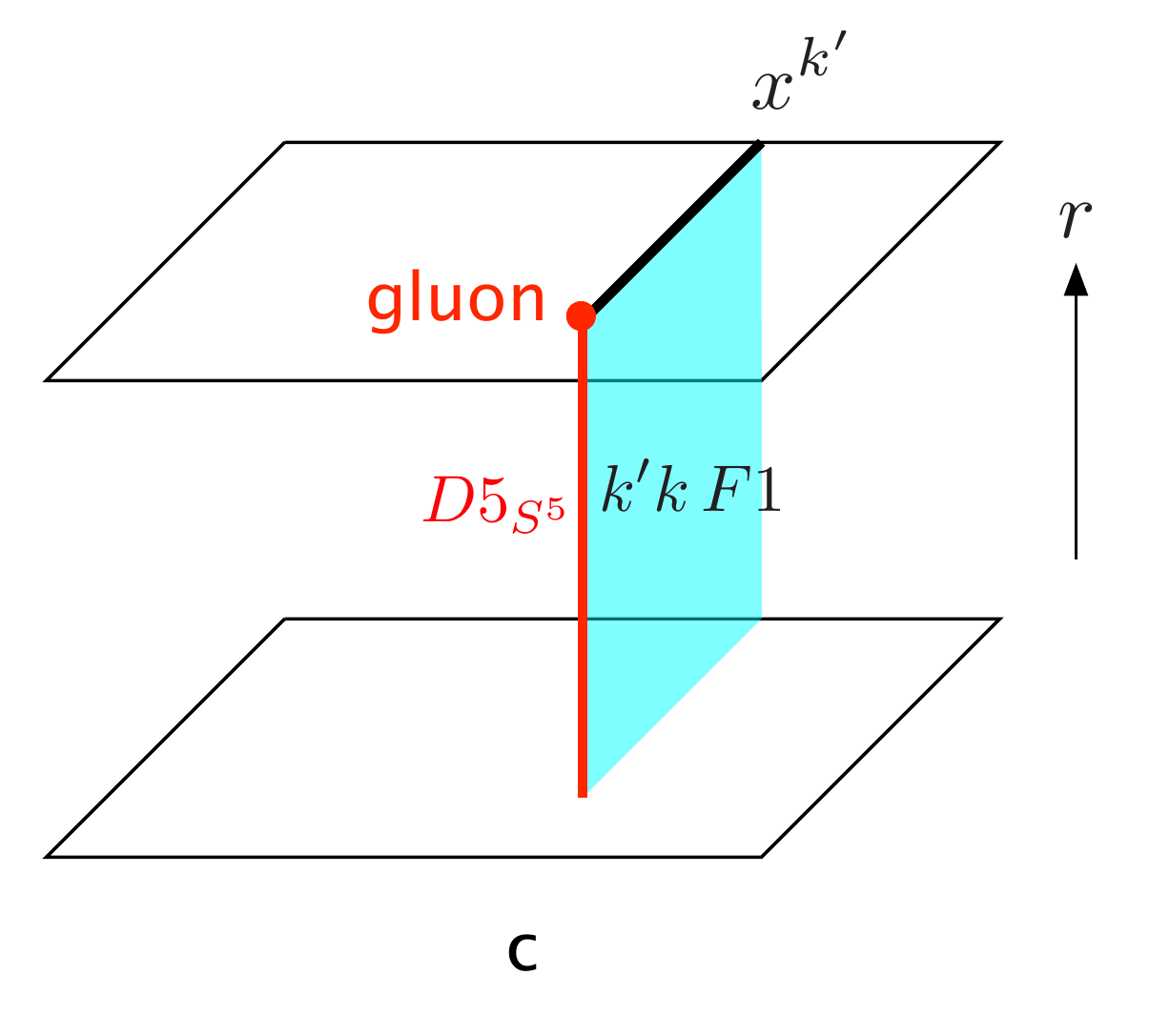} 
\caption{Bulk description of electric line operators in the $(SU(N)/\mathbb{Z}_k)_\ell$ theory: (a) $n=k$ is a genuine line,
 (b) lines with $n<k$ are not genuine, (c) $k'$ genuine lines are screened by a gluon.}
\label{WilsonLinesSU(N)/Z_k}
\end{figure}

\begin{figure}[h!]
\center
\includegraphics[height=0.2\textwidth]{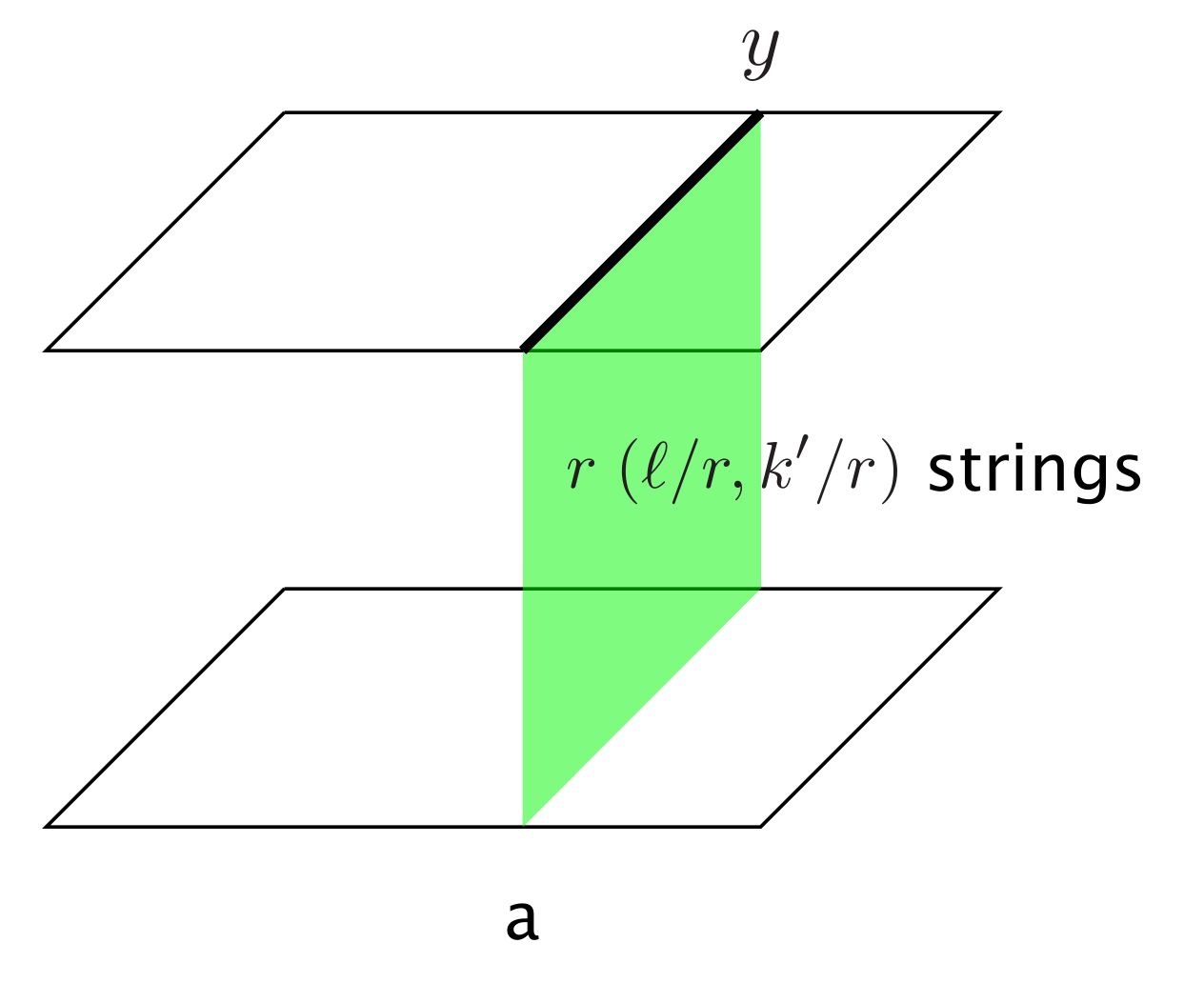} 
\hspace{10pt}
\includegraphics[height=0.2\textwidth]{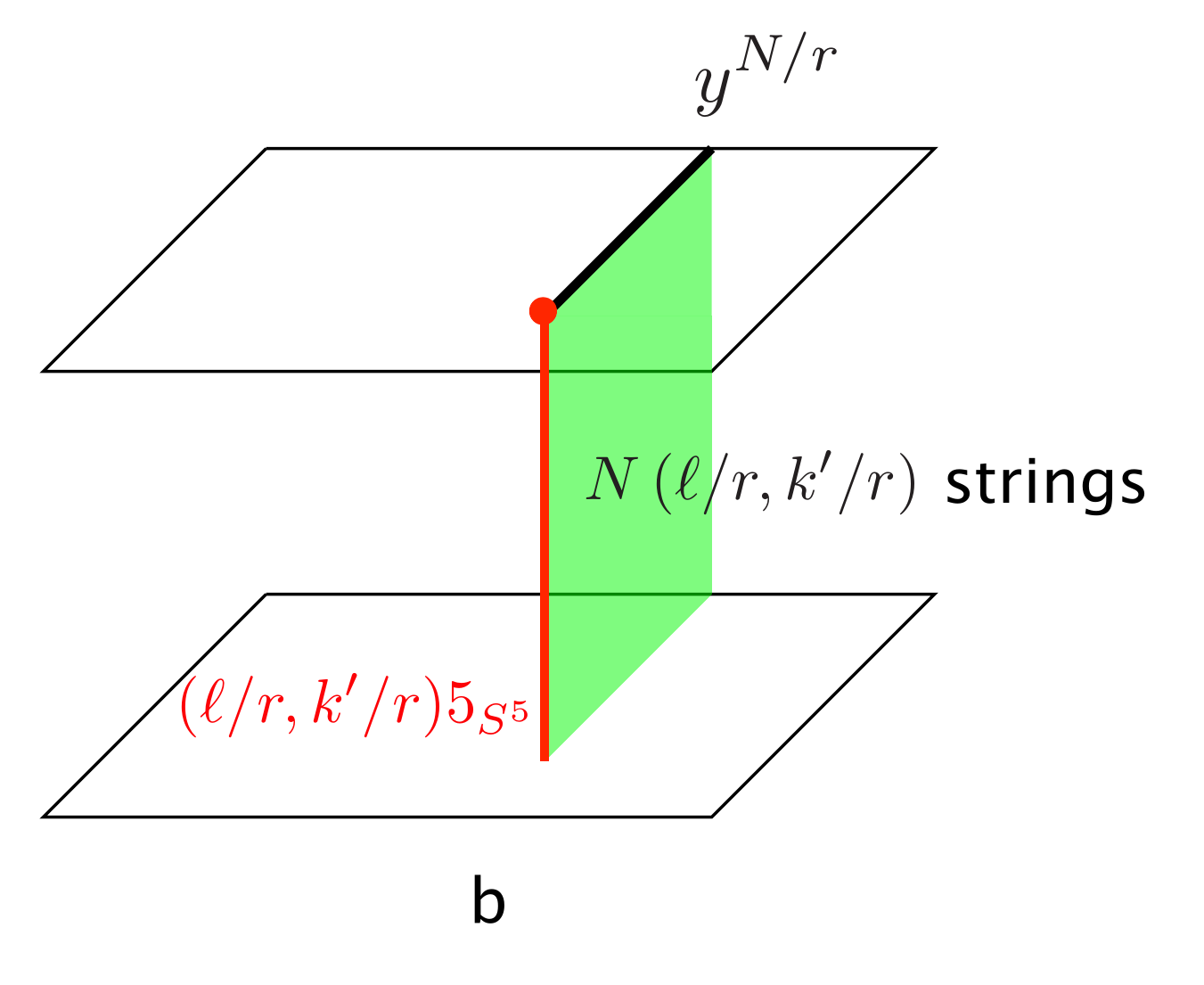}
\hspace{10pt} 
\includegraphics[height=0.2\textwidth]{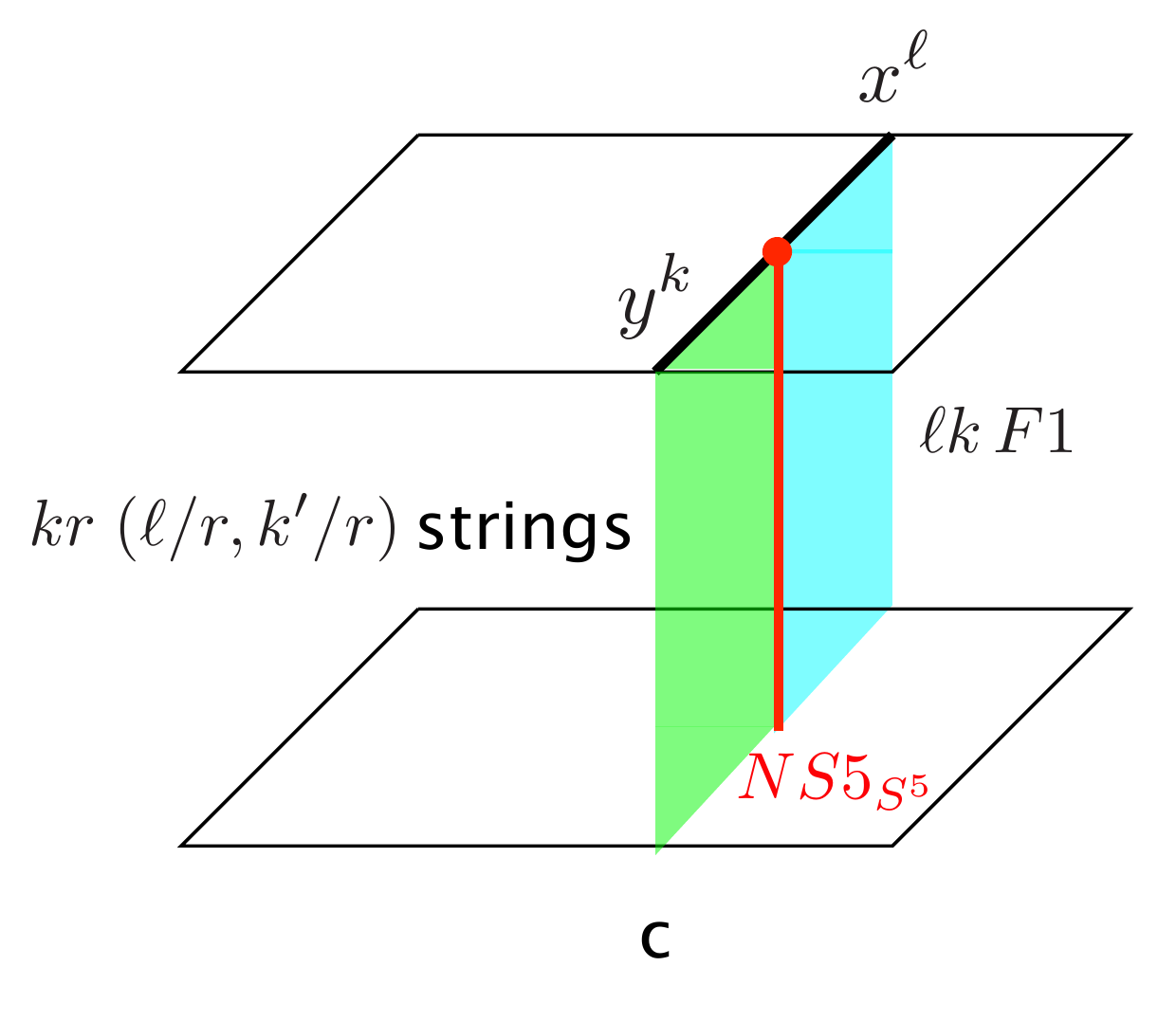} 
\caption{Bulk description of dyonic line operators in the $(SU(N)/\mathbb{Z}_k)_\ell$ theory: (a) $r$ $(\ell/r,k'/r)$ strings describe a genuine dyonic line operator ($r\equiv \mbox{gcd}(k',\ell)$).  
(b) $N/r$ dyonic lines are screened by an $(\ell/r,k'/r)$5-brane on $S^5$. (c) $k$ dyonic lines can turn into $\ell$ electric lines via an NS5-brane on $S^5$.}
\label{DyonLinesSU(N)/Z_k}
\end{figure}

More generally, the first boundary condition in (\ref{SUBC}) fixing $kb$ implies that a multiple of $k$ F-strings at the boundary corresponds to a trivial 
surface operator.
We therefore have genuine electric lines described by multiples of $k$ F-strings ending on the boundary, Fig.~\ref{WilsonLinesSU(N)/Z_k}a.
A smaller number of F-strings $n<k$ ending on the boundary can attach to F-strings at the boundary, Fig.~\ref{WilsonLinesSU(N)/Z_k}b.
A  $k'$ multiple of $k$ F-strings ending on the boundary can be screened by a D5-brane wrapping $S^5$, Fig.~\ref{WilsonLinesSU(N)/Z_k}c.
This explains the $\mathbb{Z}_{k'}$ factor in (\ref{OneFormSymmetry1}). 
The second boundary condition fixing $k'c + \ell b$ imposes an analogous constraint on $(p,q)$ strings with 
$(p,q)=(\ell,k')/\mbox{gcd}(k',\ell)$. 
Namely a multiple of $\mbox{gcd}(k',\ell)$ $(p,q)$-strings at the boundary corresponds to a trivial surface operator,
and therefore multiples of $\mbox{gcd}(k',\ell)$ such strings ending on the boundary describe genuine dyonic line operators, 
Fig.~\ref{DyonLinesSU(N)/Z_k}a.
A smaller number of $(p,q)$-strings ending on the boundary can attach to $(p,q)$ strings at the boundary.
An $N/\mbox{gcd}(k',\ell)$ multiple of $\mbox{gcd}(k',\ell)$ of these strings can be screened by an $(p,q)$ 5-brane wrapping $S^5$, Fig.~\ref{DyonLinesSU(N)/Z_k}b.
This explains the $\mathbb{Z}_{N/{gcd}(k',\ell)}$ factor in (\ref{OneFormSymmetry1}).
Finally, the divisor $\mathbb{Z}_{k'/gcd(k',\ell)}$ in (\ref{OneFormSymmetry1}) corresponds to the identification of 
$k\, \mbox{gcd}(k',\ell)$ $(p,q)$-strings
with $\ell k$ F-strings via an NS5-brane on $S^5$ (Fig.~\ref{DyonLinesSU(N)/Z_k}c).

Altogether this reproduces the spectrum of line operators in (\ref{LineOperatorSpectrum}), 
and the one form symmetry in (\ref{OneFormSymmetry1}), or equivalently (\ref{OneFormSymmetry2}).

\subsection{The interface as a two-face}

In pure $SU(N)$ YM theory, the mixed anomaly at $\theta = \pi$ implies that CP is spontaneously broken in the IR,
and therefore that there exists a domain wall separating the distinct vacuua with $\theta = \pi$ and $\theta = -\pi$ 
\cite{Gaiotto:2017tne}.\footnote{Spontaneous CP breaking is actually one of three logical possibilities for the IR theory. 
The other two are a non-trivial gapless theory or a gapped topological theory.
However it is known that CP is spontaneously broken for $N\rightarrow \infty$, and so it is plausible that this is true for finite and large $N$.}
The anomaly implies that the 3d domain wall theory is an $SU(N)_1$ CS theory.
The ${\cal N}=4$ theory, on the other hand, is a conformal field theory, so CP cannot be spontaneously broken, and there is no domain wall.
However one can still have a nontrivial interface separating two regions with values of $\theta$ differing by a multiple of $2\pi$,
say $\theta = 0$ and $\theta = 2\pi k$.
In the pure YM theory the 3d interface theory depends on the magnitude of the derivative of $\theta$ relative to 
the dynamical scale of the theory $\Lambda$ \cite{Gaiotto:2017tne}.
For $|\nabla \theta| \ll \Lambda$ the interface corresponds essentially to $k$ domain walls, and the interface theory is $(SU(N)_1))^k$.
For $|\nabla \theta| \gg \Lambda$ the interface theory is $SU(N)_k$.
There must therefore be a transition between these two theories at some intermediate scale.
In contrast, the ${\cal N}=4$ theory does not have a dynamical scale, and the interface theory is just $SU(N)_k$.
Using level-rank duality we can equivalently describe the interface theory as $U(k)_N$.

The holographic dual of an interface should involve a D7-brane, which sources the Type IIB axion $C_0$ dual to
the $\theta$ parameter of the SYM theory.
In the original brane configuration in flat space we have:
\medskip
\begin{center}
\begin{tabular}{ |c|c|c|c|c|c|c|c|c|c|c| } 
\hline
 & 0 & 1 & 2 & 3 & 4 & 5 & 6 & 7 & 8 & 9 \\ 
 \hline
 D3 & $\times$ & $\times$ & $\times$ & $\times$ &   &   &   &   &   &   \\ 
 \hline
 D7 & $\times$ & $\times$ & $\times$ &   &   & $\times$ & $\times$ & $\times$ & $\times$ & $\times$\\ 
 \hline
\end{tabular}
\end{center}
This is an $ND=6$ brane system and therefore non-supersymmetric.
At non-zero separation along $x_4$, the D3-branes and D7-branes repel each other.
Replacing the $N$ D3-branes with their $AdS_5\times S^5$ near-horizon background,
these D7-branes wrap the $S^5$ and are transverse to the $x_3$ and radial coordinates of $AdS_5$.
The D7-branes are pushed towards the $AdS_5$ horizon.

A related configuration was studied in \cite{Fujita:2009kw}. 
There the $x_3$ coordinate was compactified such that $x_3 \sim x_3 + L$, leading to a
background, known as the $AdS_5$ soliton, in which the $x_3$ circle shrinks at a non-zero radial position $r_0 = \frac{\pi R^2}{L}$.
The $(x_3,r)$ space is then topologically a disk with the D7-branes located at its center.
Taking $k$ D7-branes, and ignoring their backreaction on the metric and dilaton, the axion field takes the form
\be 
C_0(x_3) = \frac{k}{L} x_3 \,.
\ee
From the point of view of the boundary theory, namely the low-energy 3d worldvolume theory of the $N$ compactified D3-branes,
this gives a level $k$ $SU(N)$ CS term.
On the other hand, from the point of the D7-branes, the 5-form flux on $S^5$ gives a level $N$ $U(k)$ CS term.
Thus this construction gives a holographic realization of level-rank duality.

In our case $x_3$ is not compact, and the dependence of the axion on $x_3$ and $r$ is more complicated.
But in principle this should provide a realization of an interface that interpolates between $\theta(x_3\rightarrow -\infty) = 0$
and $\theta(x_3\rightarrow +\infty) = 2\pi k$.
Remarkably, there exists a fully backreacted solution with precisely this property.
This is the axionic Janus (the Roman two-faced god) 
solution, which we now describe.

\begin{figure}[h!]
\centering \includegraphics[height=2.4in]{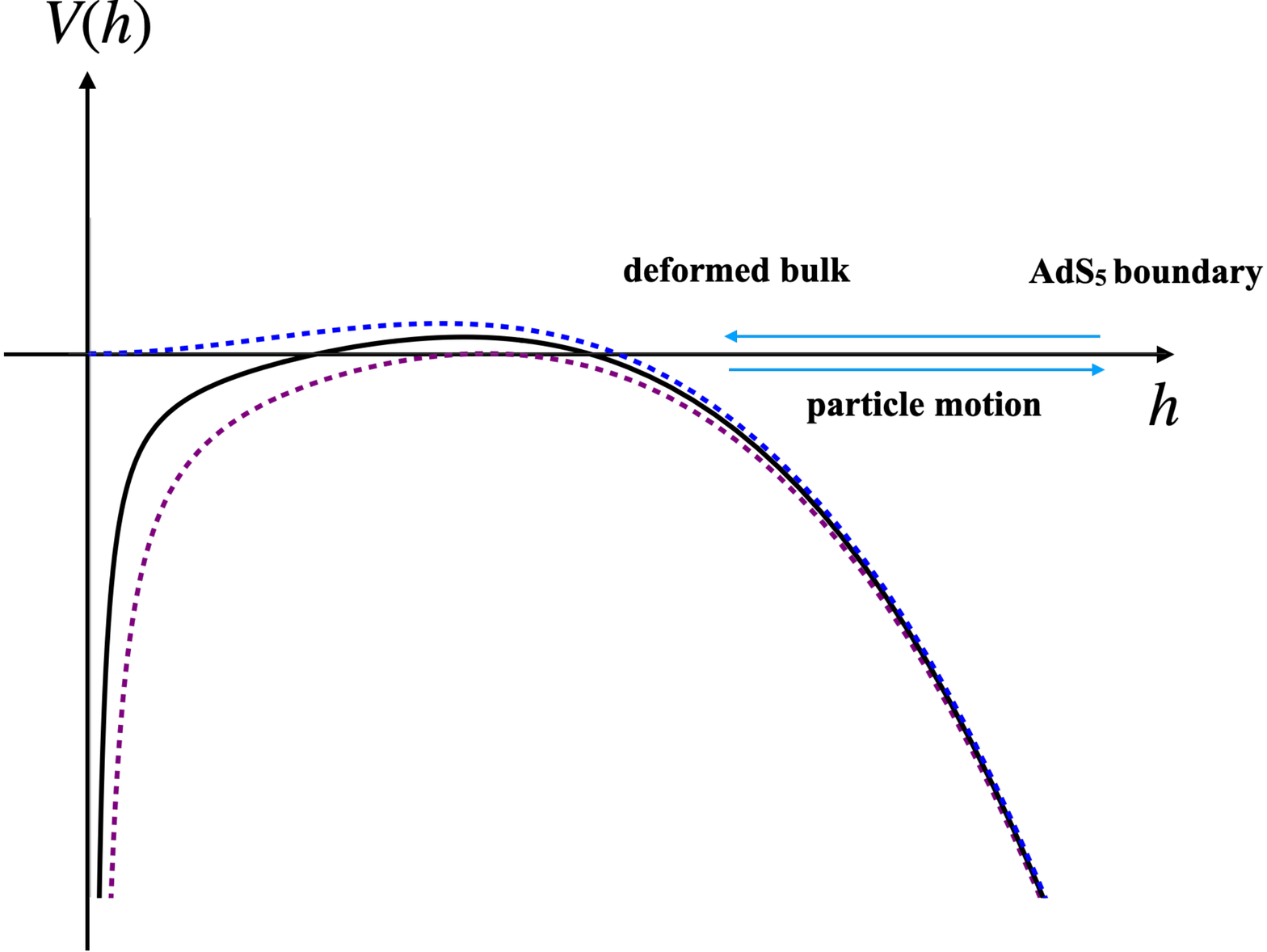}
\caption{The critical case is indicated by the dotted purple curve. The dotted blue curve is the undeformed AdS$_5$ and the solid black curve is a generic case in the range $0<c_0<c_{\ast}$.}
\label{fig:warpfactpr}
\end{figure}

The Janus geometry is given by a specific $SO(2,3)$ preserving deformation of $AdS_5\times S^5$ most easily expressed using
the $AdS_4$ slicing of $AdS_5$
\cite{Bak:2003jk,DHoker:2006vfr},
\be
\label{Janusansatz}
ds^2 &=& h(\mu)\left(d\mu^2+ds_{AdS_4}^2\right)+d\Omega_5^2 \\[5pt]
F_5 &=& 2h(\mu)^{5/2}d\mu\wedge \omega_{AdS_4}+2\omega_{S^5} \,.
\ee
For $AdS_5$, the coordinate $\mu$ is related to the usual $AdS_5$ coordinates as
\be
\label{mucoordinate}
\cos^2\mu = \frac{z^2}{z^2 + x_3^2} \,,
\ee
and it ranges from $-\pi/2$, corresponding to one half of the $z=0$ boundary with $x_3<0$, to $\pi/2$,
corresponding to the other half of the boundary with $x_3>0$. The warp factor for $AdS_5$ is given by
$h(\mu) = 1/\cos^2\mu$, and the axio-dilaton $\tau \equiv C_0 + ie^{-\phi}$ is a constant.
More generally, the reduced equation for the axio-dilaton is given by
\be
\label{tauEquation}
{\tau''\over\tau'}+{3h'\over 2h}+i{\tau'\over{\mbox{Im}(\tau)}} &=& 0 \,.
\ee
Integrating the real part of (\ref{tauEquation}) gives
\be
{|\tau'|^2\over{(\mbox{Im}(\tau)})^2}={c_0^2\over h^3} \,,
\ee
which can then be used to integrate the equations for the warp factor to give
\be 
h'^2 - 4h^3+ 4h^2 &=& {c_0^2\over 6h} \,.
\label{eq:warpfactor}
\ee
This equation can be viewed as the integrated equation of motion of a zero energy particle in a 
potential $V(h)=-4h^3+4h^2-{c_0^2\over 6h}$ with $h\in (0,\infty)$ (see Fig.~\ref{fig:warpfactpr}).
The particle comes in from infinity, corresponding to one half of the boundary at $\mu=-\mu_0$, and bounces back (at the greatest root of $V(h)=0$) to infinity, corresponding to the other half at $\mu=+\mu_0$. 
The case $c_0=0$ corresponds to the undeformed $AdS_5$ solution, for which $\mu_0 = \pi/2$.
More generally $\mu_0 > \pi/2$.
There is a critical value of $c_0^2$, given by $c^2_{\ast}=81/32$, above which the particle does not bounce back and reaches the singularity at $h=0$. 

\begin{figure}[h!]
\centering
\centering \includegraphics[height=2.1in]{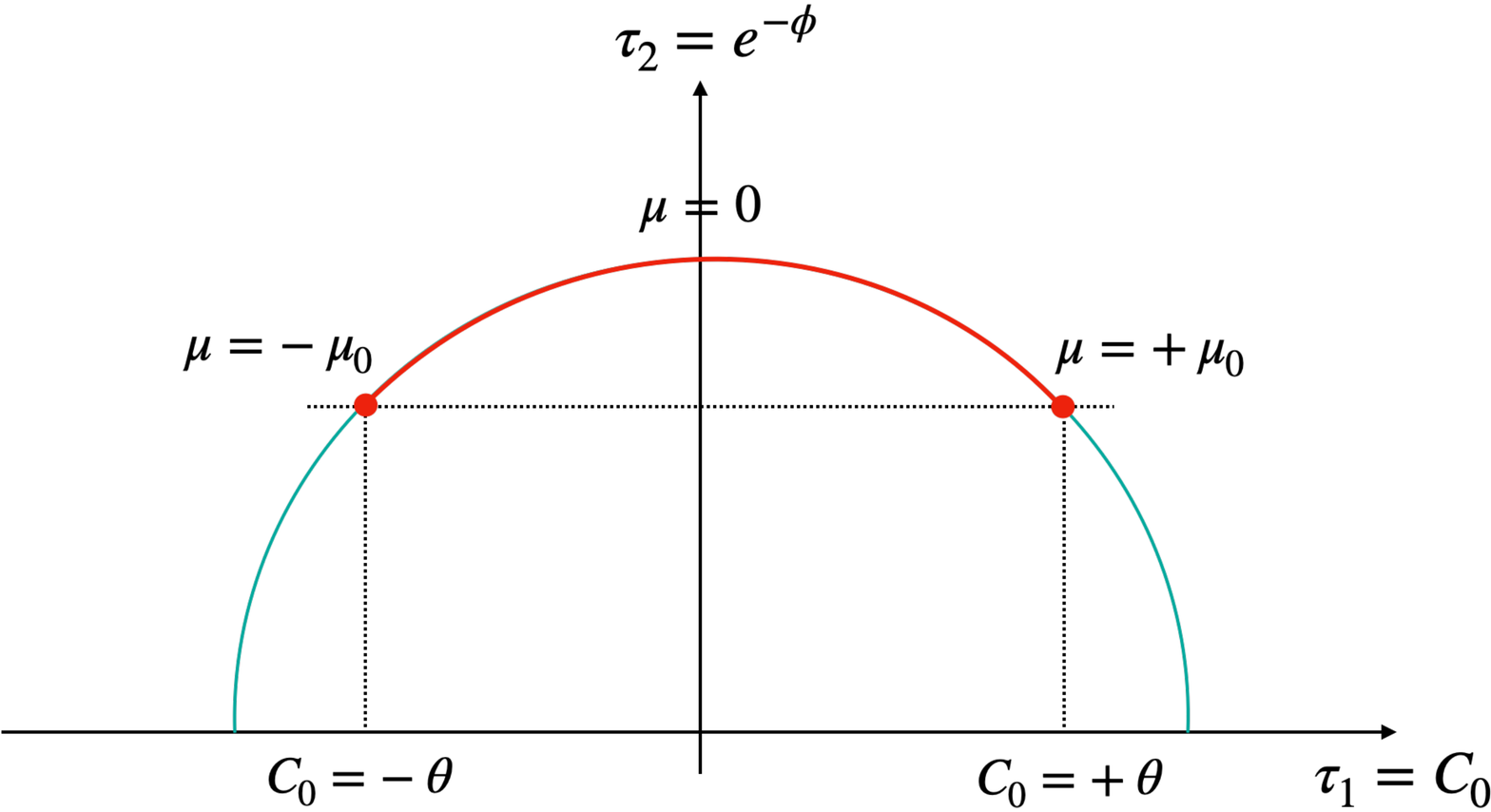}
\caption{The profile of the axio-dilaton: The axion $C_0=\tau_1$ varies from $-\theta$ at one half of the boundary to $+\theta$ at the other half of the boundary. In the boundary, the axion $C_0$ jumps across the domain wall. }
\label{fig:AxioDilaton}
\end{figure}

Integrating the imaginary part of (\ref{tauEquation}) gives\footnote{This only holds if $\mbox{Re}(\tau)$ is not a constant.
If $\mbox{Re}(\tau)$ is a constant, the imaginary part of (\ref{tauEquation}) is trivially satisfied, and the solution reduces to the 
purely dilatonic Janus solution of \cite{Bak:2003jk}.}
\be
|\tau(\mu) - p|^2 = r^2 \,, \qquad p,r\in\mathbb{R} \,.
\ee
In other words the axio-dilaton resides on a semicircle of radius $r$ in the upper half complex plain. 
The general solution interpolates between a particular value of $\tau$ on this circle on one half of the boundary at $\mu = -\mu_0$ and another value on the other half at $\mu = \mu_0$.
Of particular interest to us are trajectories such that the dilaton is the same on both halves of the boundary.
For example, if we set $p=0$, the solution will interpolate between $(C_0,e^{-\phi})(-\mu_0) = (-\theta,\sqrt{r^2-\theta^2})$ and 
$(C_0,e^{-\phi})(\mu_0) = (\theta,\sqrt{r^2-\theta^2})$, Fig.~\ref{fig:AxioDilaton}.
Note that the dilaton remains small everywhere as long as $r^2 - \theta^2 \gg 1$.

\begin{figure}[h!]
\centering
\centering \includegraphics[height=1.8in]{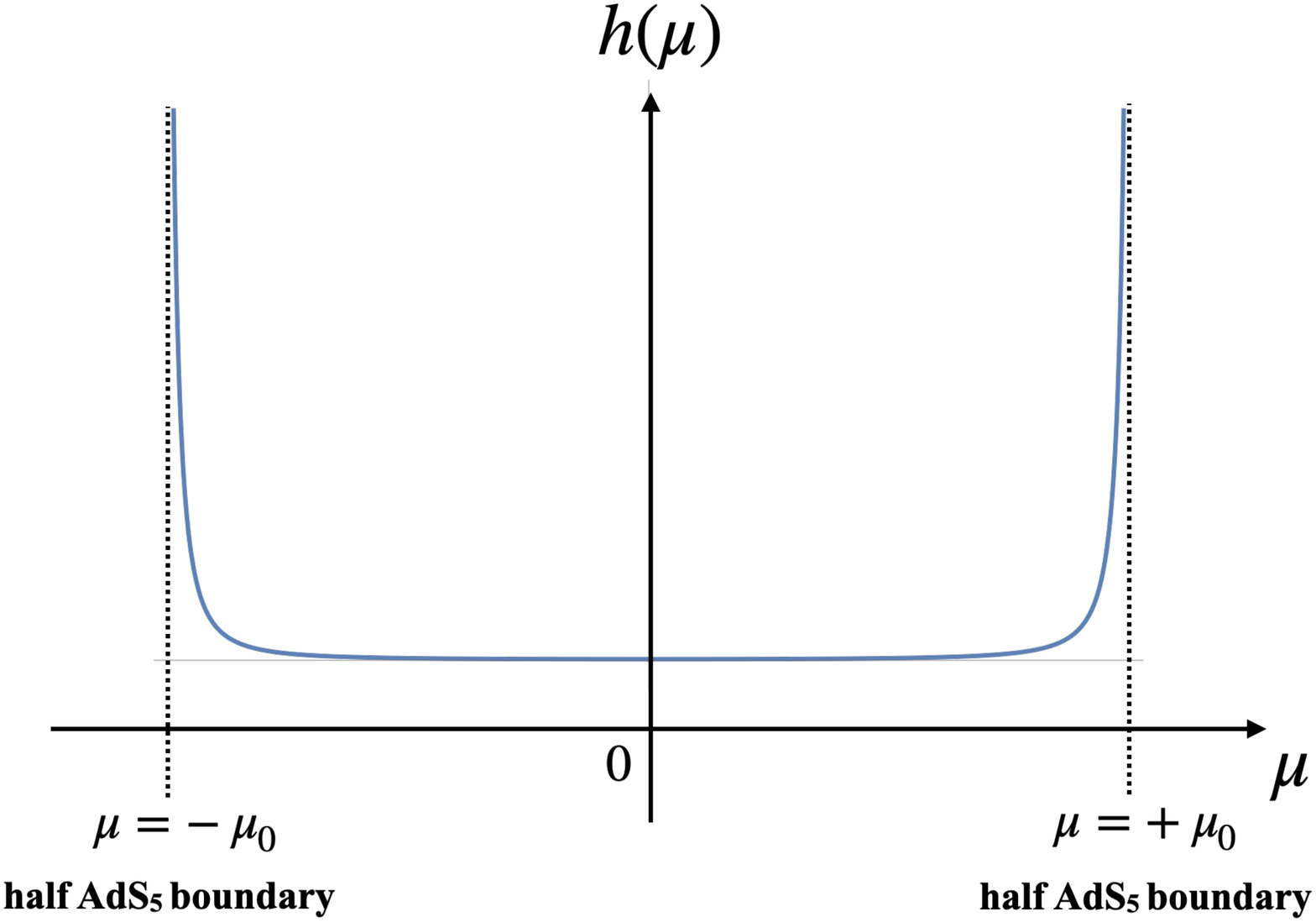}
\hspace{1.8cm}
\centering \includegraphics[height=1.8in]{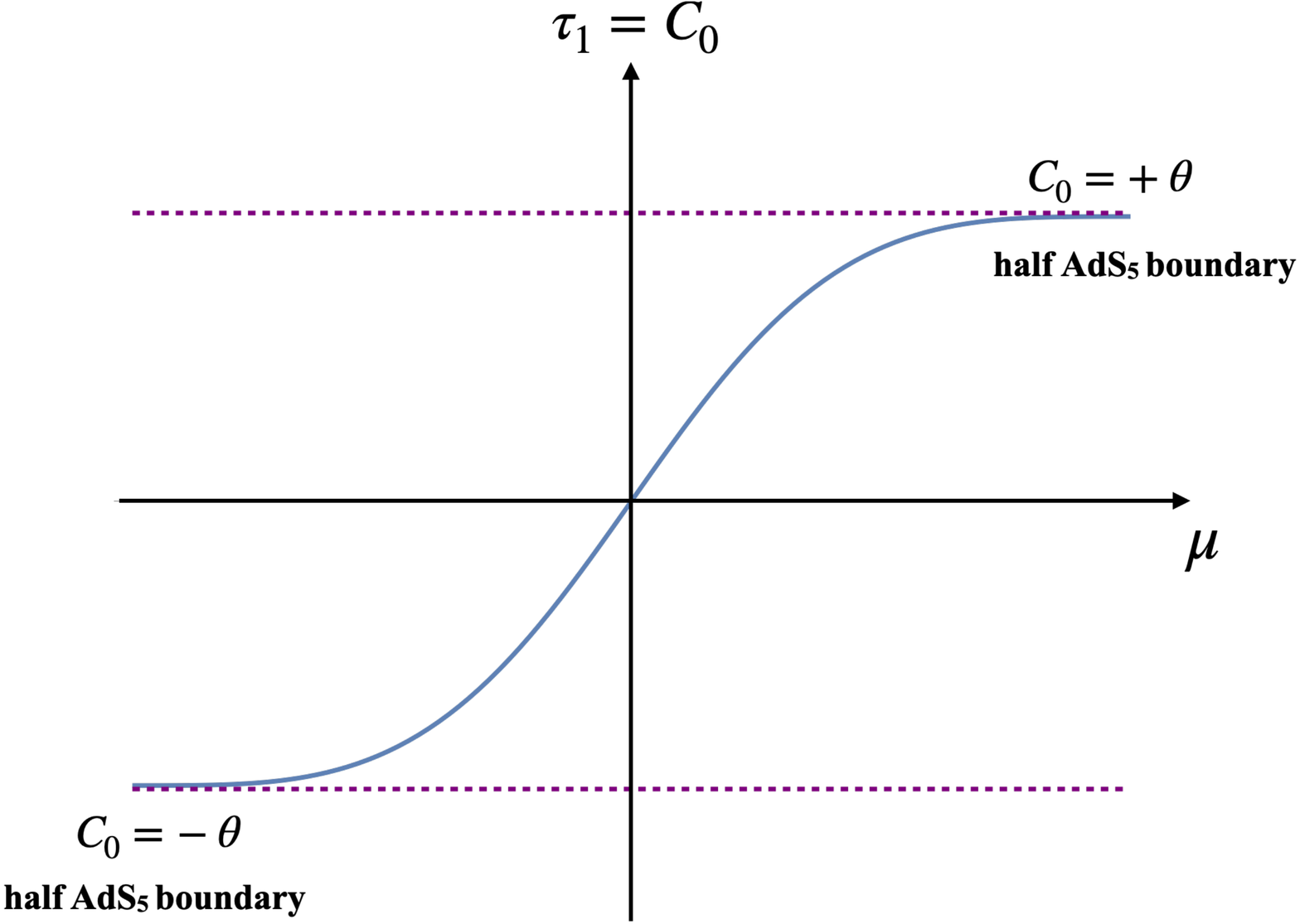}
\caption{(Left) The warp factor $h(\mu)$. Each end where $h(\mu)$ brows up corresponds to one half of the boundary space (at $\mu=\pm\mu_0$) separated by an interface. (Right) The axion profile $C_0=\tau_1(\mu)$. The axion varies from $C_0=-\theta$ to $C_0=+\theta$ as it goes from one half of the boundary to the other through the bulk.}
\label{fig:Anus}
\end{figure}

The analytic solution is known and can be expressed in terms of elliptic functions \cite{DHoker:2006vfr}.
In Fig.~\ref{fig:Anus}  we present numerical plots of the warp factor $h(\mu)$ and the axion $C_0(\mu)$.
Note that the axion is a smooth function of $\mu$, but is a step function as a function of $x_3$ on the boundary.
The coordinate relation in $AdS_5$ (\ref{mucoordinate}) can be rewritten as
\be
\mu = \mbox{} \frac{i}{2}\mbox{ln} \frac{x_3-iz}{x_3+iz} -\frac{\pi}{2} \,.
\ee
At the boundary $z=0$ this becomes 
\be
\mu \stackrel{z\rightarrow 0}{\longrightarrow} \pi \left(\Theta(x_3) - \frac{1}{2}\right)
\ee
Since the Janus geometry is asymptotically $AdS_5$, the coordinate relation is essentially the same for $z\ll1$,
modulo a rescaling of $\mu$, and we have
\be
\mu \stackrel{z\rightarrow 0}{\longrightarrow} 2\mu_0 \left(\Theta(x_3) - \frac{1}{2}\right) \,.
\ee
Charge quantization requires $C_0(\mu_0) - C_0(-\mu_0) = 2\theta = 2\pi k$, where $k$ is the number of D7-branes sourcing the Janus deformation.

\section{The $so(2n)$ theories} 

\subsection{One form symmetries and dualities}

The different theories are obtained by gauging subgroups of the one-form symmetry of the covering group $Spin(2n)$.
The group $Spin(2n)$ has four classes of representations: the adjoint, or trivial, class $I$, the vector class $V$, and the two
spinor classes $S$ and $C$.

\begin{figure}[h!]
\center
\includegraphics[height=0.3\textwidth]{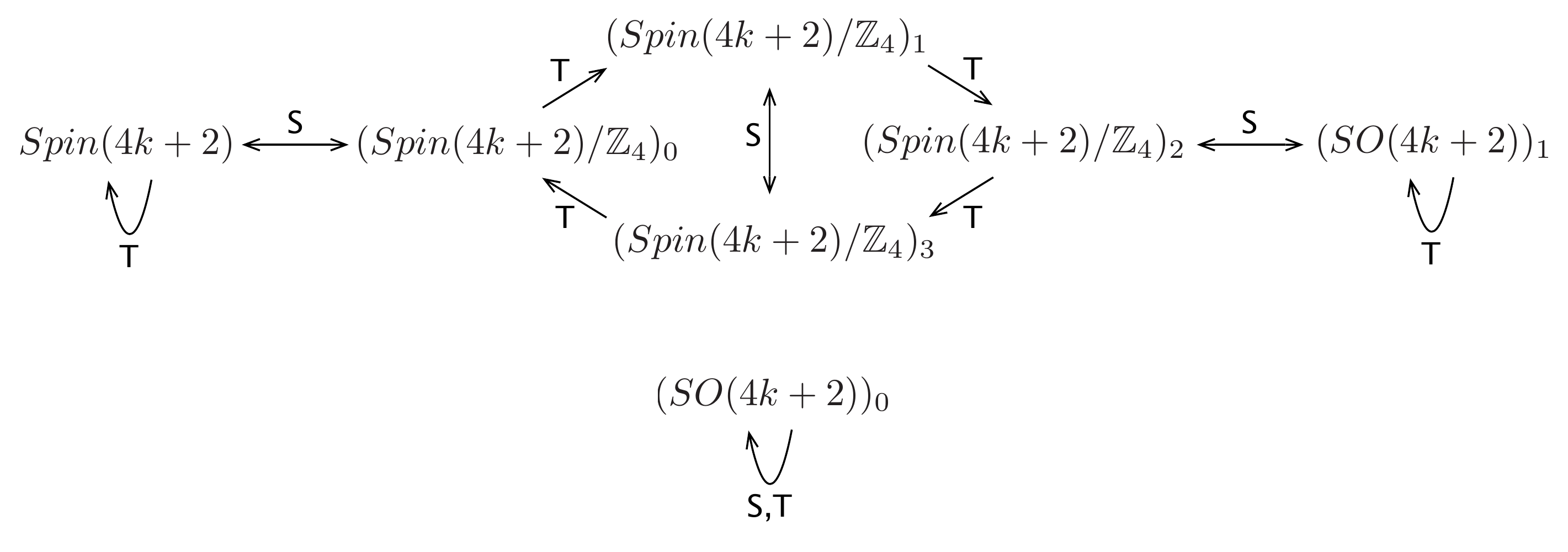} 
\caption{The $SL(2,\mathbb{Z})$ duality orbits for $so(4k+2)$, reproduced from \cite{Aharony:2013hda}.}
\label{so(4k+2)Duality}
\end{figure}

For odd $n$ the center of $Spin(2n)$ is $\mathbb{Z}_4$, with the different classes of representations transforming under the generator as
\be
I \rightarrow  I \; , \;
S  \rightarrow  iS \; ,\;
C  \rightarrow  -iC \; , \;
V  \rightarrow  -V \,.
\ee
In particular for odd $n$ we have the relations $S\times S = C\times C = V$, $S\times C = I$, $S\times V = C$, and $C\times V=S$.
The possible gauge groups are $Spin(4k+2)$,
$Spin(4k+2)/\mathbb{Z}_2 = SO(4k+2)$, and $Spin(4k+2)/\mathbb{Z}_4$.
The line operator charges $z_e,z_m$  take values in $\mathbb{Z}_4$, with the generator corresponding to $S$.
The Dirac pairing condition is given by
\be
z_e z_m' - z_m z_e' = 0 \; \mbox{mod} \; 4 \,.
\ee
There are seven different maximal charge lattices satisfying this condition.
The corresponding theories and their one-form symmetries are shown in Table~\ref{Table1}.\footnote{In \cite{Aharony:2013hda}
the additional parameter for $SO(2n)$ was denoted as $\pm$.}
Note that this generalizes $so(6)=su(4)$ from the previous section.
The action of $SL(2,\mathbb{Z})$ is shown in Fig.~\ref{so(4k+2)Duality}.
There are two orbits of theories for any $k$, one containing the six theories with $G^{(1)}=\mathbb{Z}_4$,
and one with just the $SO(4k+2)_0$ theory which has $G^{(1)}=\mathbb{Z}_2 \times \mathbb{Z}_2$.
\begin{table}[h!]
\begin{center}
\begin{tabular}{|l|l|l|}
  \hline 
  theory & $(z_e,z_m)$  & $G^{(1)}$ \\
 \hline
  $Spin(4k+2)$ & $(S,I)^n$  & $\mathbb{Z}_4$ \\
   \hline
  $SO(4k+2)_0$ & $(V^n,V^m)$ & $\mathbb{Z}_2 \times \mathbb{Z}_2$ \\
  \hline
  $SO(4k+2)_1$ & $(S,V)^n$ & $\mathbb{Z}_4$ \\
   \hline
   $(Spin(4k+2)/\mathbb{Z}_4)_{\ell}$ & $(S^{\ell},S)^n$ &  $\mathbb{Z}_4$ \\
  \hline
    \end{tabular}
 \end{center}
\caption{The seven $so(4k+2)$ theories.}
\label{Table1}
\end{table}

\begin{figure}[h!]
\center
\includegraphics[height=0.5\textwidth]{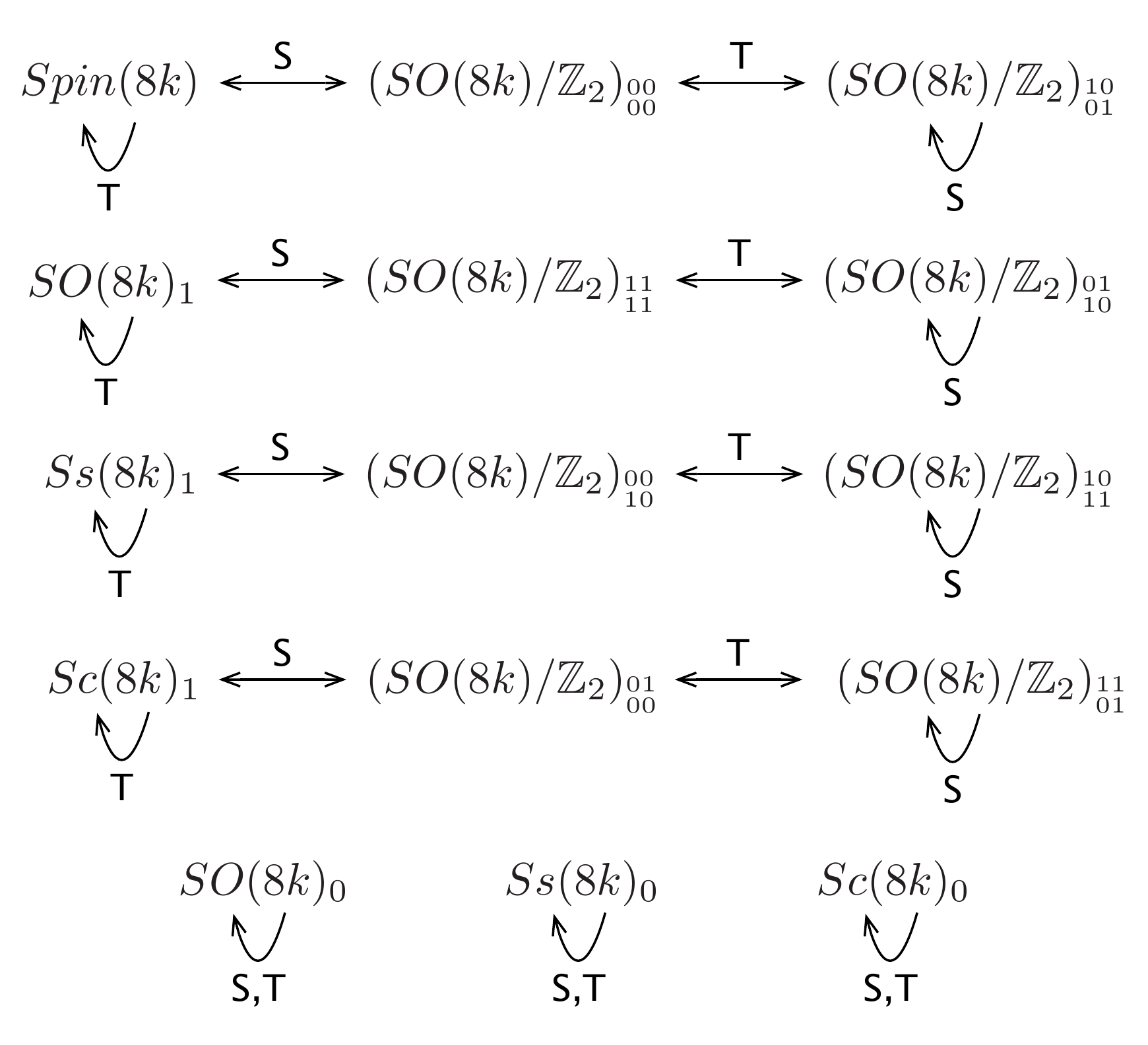} 
\caption{The $SL(2,\mathbb{Z})$ duality orbits for $so(8k)$, reproduced from \cite{Aharony:2013hda}.}
\label{so(8k)Duality}
\end{figure}

\begin{figure}[h!]
\center
\includegraphics[height=0.5\textwidth]{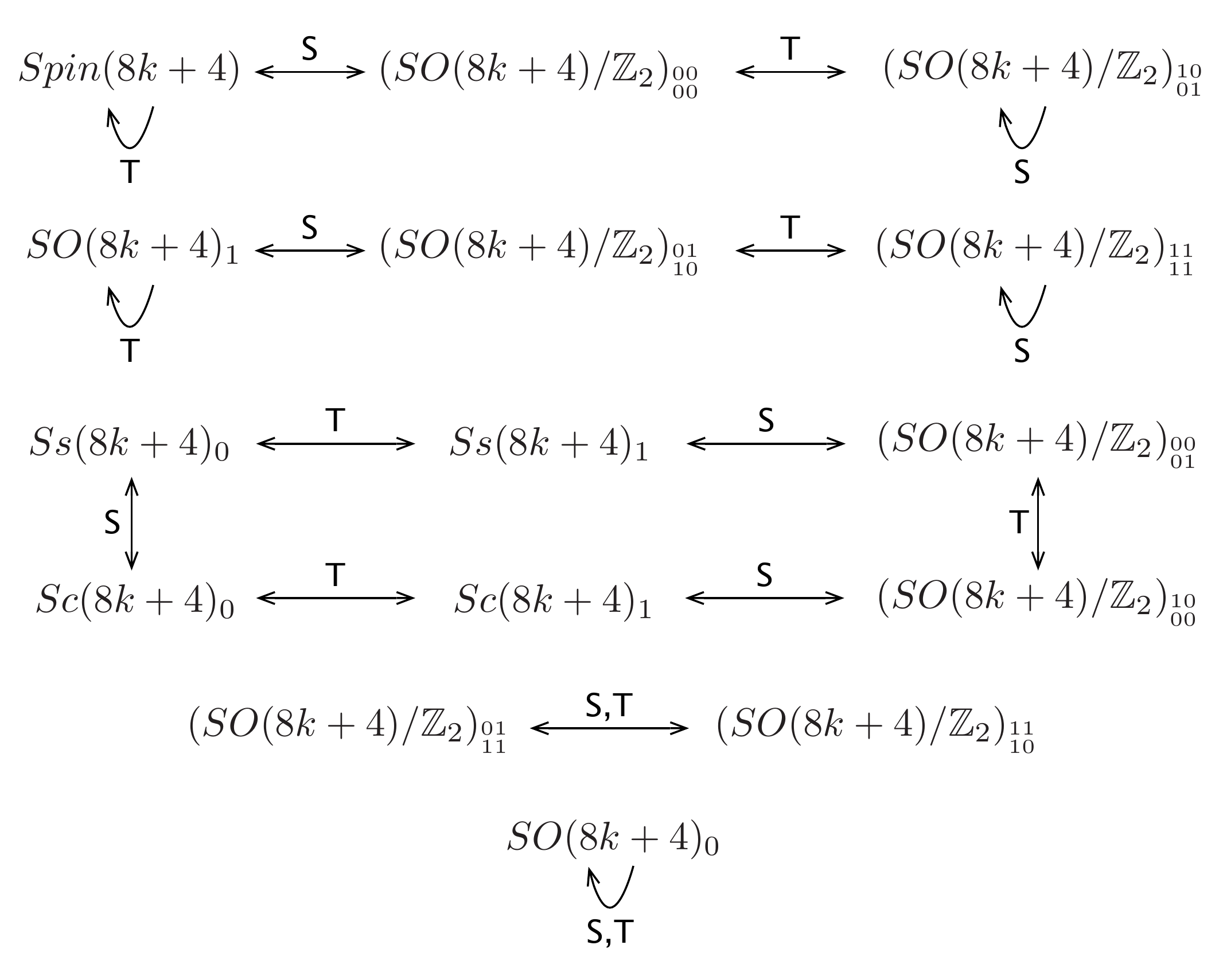} 
\caption{The $SL(2,\mathbb{Z})$ duality orbits for $so(8k+4)$, reproduced from \cite{Aharony:2013hda}.}
\label{so(8k+4)Duality}
\end{figure}

For even $n$ the center of $Spin(2n)$ is $\mathbb{Z}_2^S\times \mathbb{Z}_2^C$, with the different classes transforming as
\be
I \rightarrow  I \; , \;
S  \rightarrow  -S \; ,\;
C  \rightarrow  C \; , \;
V  \rightarrow  -V \,,
\ee
under the generator of $\mathbb{Z}_2^S$, and as
\be
I \rightarrow  I \; , \;
S  \rightarrow  S \; ,\;
C  \rightarrow  -C \; , \;
V  \rightarrow  -V \,,
\ee
under the generator of $\mathbb{Z}_2^C$.
In this case the classes are related as $S\times C = V$, $S\times S = C\times C = I$, $S\times V=C$, and $C\times V=S$.
The possible gauge groups are $Spin(2n)$,
$Spin(2n)/\mathbb{Z}_2^V = SO(2n)$ (where $\mathbb{Z}_2^V$ is the diagonal subgroup of $\mathbb{Z}_2^S\times \mathbb{Z}_2^C$), 
$Spin(2n)/\mathbb{Z}^S_2 = Ss(2n)$, $Spin(2n)/\mathbb{Z}^C_2=Sc(2n)$,
and $Spin(2n)/(\mathbb{Z}_2 \times \mathbb{Z}_2) = SO(2n)/\mathbb{Z}_2$.
The line operator charges now take values in $\mathbb{Z}_2^S\times \mathbb{Z}_2^C$, with the generator of $\mathbb{Z}_2^S$
corresponding to the spinor class $S$, and the generator of $\mathbb{Z}_2^C$
corresponding to the spinor class $C$.
The mutual locality condition is now given by \cite{Aharony:2013hda}
\be
\label{so(4k)Locality}
z_{e,S} z_{m,S}' - z_{m,S}z_{e,S}' + z_{e,C}z_{m,C}' - z_{m,C}z_{e,C}' & = & 0 \; \mbox{mod} \; 2 \quad (2n = 8k+4) \\
z_{e,S} z_{m,C}' - z_{m,C}z_{e,S}' + z_{e,C}z_{m,S}' - z_{m,S}z_{e,C}' & = & 0 \; \mbox{mod} \; 2 \quad (2n = 8k) \,.
\ee
In either case there are fifteen maximal charge lattices, all two-dimensional, as summarized in 
Table~\ref{Table2}.\footnote{For a proof see Appendix B.}
The action of $SL(2,\mathbb{Z})$ in the two cases is shown in Figs.~\ref{so(8k)Duality} and \ref{so(8k+4)Duality}.

\begin{table}[h!]
\begin{center}
\begin{tabular}{|l|l|l|}
  \hline 
  theory & $(z_{e_S},z_{e_C},z_{m_S},z_{m_C})$ & $G^{(1)}$ \\
 \hline
  $Spin(8k+4j)$ & $(S^n,C^m,I,I)$ & $\mathbb{Z}_2\times \mathbb{Z}_2$ \\
\hline
   $SO(8k+4j)_{\ell_V}$ & $(S^{n+\ell_V m},C^n,S^m,C^m)$ &  $\mathbb{Z}_2\times \mathbb{Z}_2$\\
  \hline
  $Ss(8k+4j)_{\ell_S}$ & $(S^n,C^{\ell_S m},S^{(j+1)m},C^{jm})$ &  $\mathbb{Z}_2\times \mathbb{Z}_2$\\
 \hline
 $Sc(8k+4j)_{\ell_C}$ & $(S^{\ell_C m},C^n,S^{jm},C^{(j+1)m})$ &  $\mathbb{Z}_2\times \mathbb{Z}_2$\\
 \hline
  $(SO(8k+4j)/\mathbb{Z}_2)_{^{\ell_{SS}\ell_{SC}}_{\ell_{CS}\ell_{CC}}}$ & $(S^{\ell_{SS}n+\ell_{CS}m},C^{\ell_{SC}n+\ell_{CC}m},S^n,C^m)$ 
 &  $\mathbb{Z}_2\times \mathbb{Z}_2$\\[6pt]
  \hline
 \end{tabular}
 \end{center}
\caption{The fifteen $so(8k+4j)$ theories. In the last class of theories mutual locality requires $\ell_{SS}=\ell_{CC}$ for $j=0$, and $\ell_{SC}=\ell_{CS}$ for 
$j=1$.}
\label{Table2}
\end{table}

\subsection{Holography}

The holographic dual of the $so(2n)$ theories is $AdS_5\times \mathbb{R}P^5$,
corresponding to the near horizon background of $n$ D3-branes on an orientifold 3-plane $O3^-$ \cite{Witten:1998xy}.
The orientifold action relates antipodal points on the $S^5$, and at the same time acts as $-1$ on both the NSNS and RR 2-form
potentials $B_2$, $C_2$.
The corresponding strings are therefore $\mathbb{Z}_2$ charged.
There are two additional 2-form potentials coming from the reduction of the 6-forms, dual to the NSNS and RR 2-forms in ten dimensions,
on the 4-cycle $\mathbb{R}P^4 \subset \mathbb{R}P^5$:
\be
\tilde{B}_2  =  \int_{\mathbb{R}P^4} B_6 \quad , \quad \tilde{C}_2 = \int_{\mathbb{R}P^4} C_6 \,.
\ee
These survive the orientifold projection since the volume form of the $S^4\subset S^5$ is odd.
The objects that are charged under these fields are 5-branes wrapping $\mathbb{R}P^4 \subset \mathbb{R}P^5$,
which Witten called ``fat strings" in \cite{Witten:1998xy}.
A crucial observation made by Witten in \cite{Witten:1998xy} is that a pair of identical wrapped 5-branes annihilate
into $n$ mod $2$ strings.\footnote{Witten demonstrated this by constructing a five-dimensional
manifold $Z$ with two $\mathbb{R}P^4$ boundaries, such that a 5-brane on $Z$ describes the annihilation process of two 5-branes
on $\mathbb{R}P^4$. The manifold $Z$ is in fact isomorphic to the whole compact space $\mathbb{R}P^5$, and therefore the 5-brane on $Z$ has a tadpole due the RR 5-form flux that must be cancelled by attaching to it $n$ mod 2 strings.}
In other words if $n$ is even two 5-branes of the same type annihilate into nothing,
but if $n$ is odd they leave behind a string of an appropriate type.
For a pair of D5-branes this is a fundamental string, and for a pair of NS5-branes it is a D-string.
The wrapped 5-branes, or ``fat strings" correspond to line operators in spinor representations, and the strings correspond to line operators 
in the vector representation. We will discuss these identifications more below.

The dominant part of the low energy action near the boundary of $AdS_5$ now has three CS couplings:
\be
\label{CSaction2}
S_{CS}[B_2,C_2,\tilde{B}_2,\tilde{C}_2] = \int_{AdS_5} \left(\frac{n}{2\pi} B_2\wedge dC_2 
+ \frac{1}{\pi} B_2\wedge d\tilde{B}_2 + \frac{1}{\pi} C_2\wedge d\tilde{C}_2\right) \,.
\ee 
The first CS term comes from the ten-dimensional CS term, and
the two other CS terms originate from the ten-dimensional kinetic terms for $B_2$ and $C_2$.
As before, this action implies that all the potentials are flat, and characterized entirely by their holonomies $b,c,\tilde{b},\tilde{c}$.
Note that although $B_2$ and $C_2$ are odd under the orientifold action, there is a $\mathbb{Z}_2$ remnant in the holonomies $b,c$.

In the quantum theory we can choose as our position operators $b$ and $\tilde{c}$.
The corresponding conjugate momenta are given by 
\be
\pi_b =  \frac{n}{2\pi}c + \frac{1}{\pi} \tilde{b}\; , \; \pi_{\tilde{c}} = \frac{1}{\pi} c \,.
\ee
This implies the non-trivial commutation relations
\be
\label{bb'commutator}
[c,\tilde{c}] = [b,\tilde{b}] = \pi i \;\; \mbox{mod} \;\; 2\pi i\,,
\ee
and
\be
\label{b'c'commutator}
[\tilde{b},\tilde{c}]=\frac{n \pi i}{2} \;\; \mbox{mod} \;\; 2\pi i \,.
\ee
All other commutators vanish mod $2\pi i$.
However due to the 5-brane annihilation process described above, we have the following constraints relating the variables,
\be
\label{HolonomyRelations}
nb= 2\tilde{c} \; , \; nc=2\tilde{b} \,.
\ee

If $n$ is odd, namely for $so(4k+2)$, the commutator in (\ref{b'c'commutator}) implies that $\tilde{b}$ and $\tilde{c}$ take values in
$\mathbb{Z}_4$. The constraints (\ref{HolonomyRelations}) imply that $b=2\tilde{c}$ and $c=2\tilde{b}$,
namely that $b$ and $c$ take values in $\mathbb{Z}_2 \subset \mathbb{Z}_4$.
This is consistent with the commutators in (\ref{bb'commutator}).
The holonomy variables $\tilde{c}, \tilde{b}$ correspond to electric and magnetic spinors, respectively,
and $b,c$ to electric and magnetic vectors, respectively.
A maximal set of commuting observables of the form $n_{\tilde{b}} \tilde{b} + n_{\tilde{c}}\tilde{c}$ 
then corresponds to a maximal lattice satisfying the condition
\be
n_{\tilde{c}} n_{\tilde{b}}' - n_{\tilde{b}} n_{\tilde{c}}' = 0 \; \mbox{mod} \; 4 \,.
\ee
So again we see that the condition of mutual commutativity of the boundary conditions corresponds to the condition
of mutual locality of the line operators.
The resulting assignment of boundary conditions to the $so(4k+2)$ theories is shown in Table~\ref{Table3}.
The action of $SL(2,\mathbb{Z})$ on the 6-form gauge fields in Type IIB string theory gives
\be
\label{IIBtrans2}
(\tilde{b},\tilde{c})  \stackrel{T}{\longrightarrow}  (\tilde{b}+\tilde{c},\tilde{c}) \; , \;
(\tilde{b},\tilde{c}) \stackrel{S}{\longrightarrow}  (-\tilde{c},\tilde{b}) \,.
\ee
This reproduces the $SL(2,\mathbb{Z})$ orbits of Fig.~\ref{so(4k+2)Duality}.

\begin{table}[h!]
\begin{center}
\begin{tabular}{|l|l|}
  \hline 
  theory & BC's \\
 \hline
  $Spin(4k+2)$ & $\tilde{c}=0$ \\
   \hline 
  $SO(4k+2)_0$ & $2\tilde{b} = 0$, $2\tilde{c}=0$\\
  \hline
  $SO(4k+2)_1$ & $\tilde{c}+2\tilde{b}=0$ \\
   \hline
   $(Spin(4k+2)/\mathbb{Z}_4)_{\ell}$ & $\tilde{b} + \ell\tilde{c} = 0$ \\
  \hline
    \end{tabular}
 \end{center}
\caption{The boundary conditions dual to the seven $so(4k+2)$ theories.}
\label{Table3}
\end{table}

If $n$ is even, namely for $so(8k+4j)$, 
the four holonomy variables $b,c,\tilde{b},\tilde{c}$ are independent, and all valued in $\mathbb{Z}_2$.
The difference between the $j=0$ case and the $j=1$ case is the $\tilde{b},\tilde{c}$ commutator (\ref{b'c'commutator}), 
which is trivial in the former case and non-trivial in the latter case.
Now we consider a set of observables of the form $n_b b + n_c c + n_{\tilde{b}}\tilde{b} + n_{\tilde{c}}\tilde{c}$.
Mutual commutativity requires that for any pair of observables we have
\be
n_b n_{\tilde{b}}' + n_{\tilde{b}} n_b' + n_c n_{\tilde{c}}' + n_{\tilde{c}} n_c' 
+ j(n_{\tilde{b}} n_{\tilde{c}}' + n_{\tilde{c}} n_{\tilde{b}}') = 0 \;\; \mbox{mod} \;\; 2 \,.
\ee
This reproduces the mutual locality conditions for the line operators in (\ref{so(4k)Locality}) once we identify
$n_{\tilde{c}} = z_{e,S}$, $n_{\tilde{b}} = z_{m,S}$, $n_b = z_{e,V} = z_{e,S} + z_{e,C}$,
and $n_c = z_{m,V} = z_{m,S} + z_{m,C}$.
The assignment of boundary conditions to the $so(8k+4j)$ theories is shown in Table~\ref{Table4}.
The duality orbits of Figs.~\ref{so(8k)Duality},\ref{so(8k+4)Duality} are reproduced by the action of $SL(2,\mathbb{Z})$
on $(b,c)$ and $(\tilde{b},\tilde{c})$.

\begin{table}[h!]
\begin{center}
\begin{tabular}{|l|l|}
  \hline 
  theory &  boundary conditions \\
 \hline
  $Spin(8k+4j)$ &  $\tilde{c}=0$  \\
  &   $b=0$ \\
  \hline
   $SO(8k+4j)_{\ell_V}$ & $b=0$ \\
   &   $c+\ell_V \tilde{c} = 0$\\
  \hline
  $Ss(8k+4j)_{\ell_S}$ & $\tilde{c}=0$ \\[3pt]
 & $\tilde{b} + jc + \ell_S b = 0$ \\ 
 \hline
 $Sc(8k+4j)_{\ell_C}$ & $\tilde{c} + b = 0$ \\[3pt]
 &  $c+\tilde{b}+ jc + \ell_C \tilde{c} = 0$  \\ 
 \hline
 $(SO(8k+4j)/\mathbb{Z}_2)_{^{\ell_{SS}\ell_{SC}}_{\ell_{CS}\ell_{CC}}}$ & $\tilde{b} + (\ell_{SS}+\ell_{SC})\tilde{c} + \ell_{SC} b =0$\\[3pt]
  & $c + \tilde{b} + (\ell_{CS}+\ell_{CC}) \tilde{c} + \ell_{CC}b =0$ \\
  \hline
 \end{tabular}
 \end{center}
\caption{The boundary conditions dual to the fifteen $so(8k+4j)$ theories. In the last class of theories we require $\ell_{SS}=\ell_{CC}$ for $j=0$, and $\ell_{SC}=\ell_{CS}$ for 
$j=1$.} 
\label{Table4}
\end{table}

\subsection{Branes and line operators}

As before, the line operators in the 4d gauge theory correspond to the boundaries of string worldsheets ending on the boundary of $AdS_5$.
Now we have both the 10d strings, namely the fundamental strings, the D-strings, and more generally $(p,q)$ strings, as well
as the various  ``fat strings" corresponding to 5-branes wrapping $\mathbb{R}P^4 \subset \mathbb{R}P^5$.
The boundaries of the fat strings correspond to line operators in one of the spinor representations of $Spin(2n)$, 
which we will take to be $S$,\footnote{This follows from the fermionic zero modes on the 5-3 strings \cite{Witten:1998xy}.}
and the boundaries of the ten-dimensional strings correspond to line operators in the vector representation $V$.
More specifically, the fundamental string corresponds to an electric vector, the D-string to a magnetic vector,
the wrapped D5-brane to an electric spinor $S$, and the wrapped NS5-brane to a magnetic spinor $S$.
A line operator in the other spinor representation $C$ is described by a bound state of a wrapped 5-brane and the appropriate string.
The full spectrum of line operators depends on the boundary conditions on the fields $b,c,\tilde{b},\tilde{c}$, since these
boundary conditions determine which strings are allowed to end on the boundary of $AdS_5$.
Let us consider just a few examples.

\medskip

\begin{figure}[h!]
\center
\includegraphics[height=0.2\textwidth]{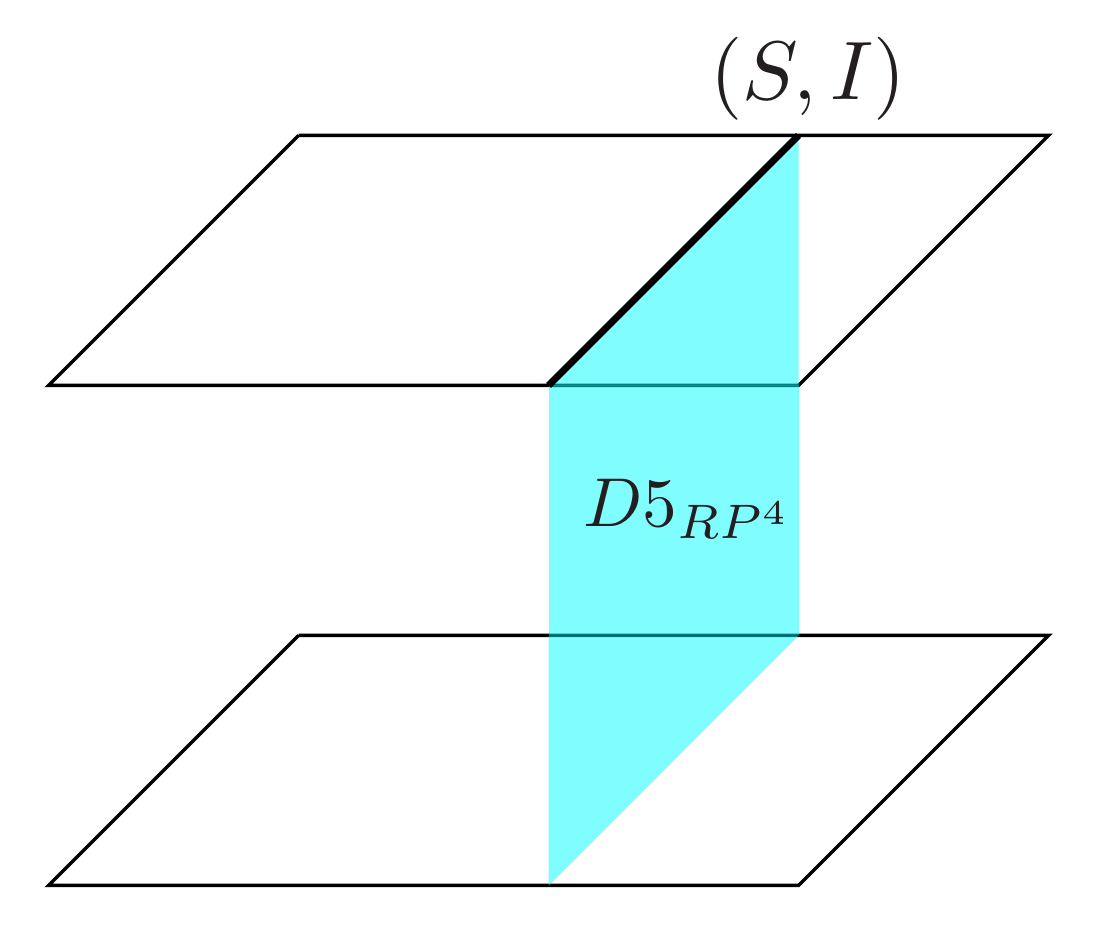}
\includegraphics[height=0.2\textwidth]{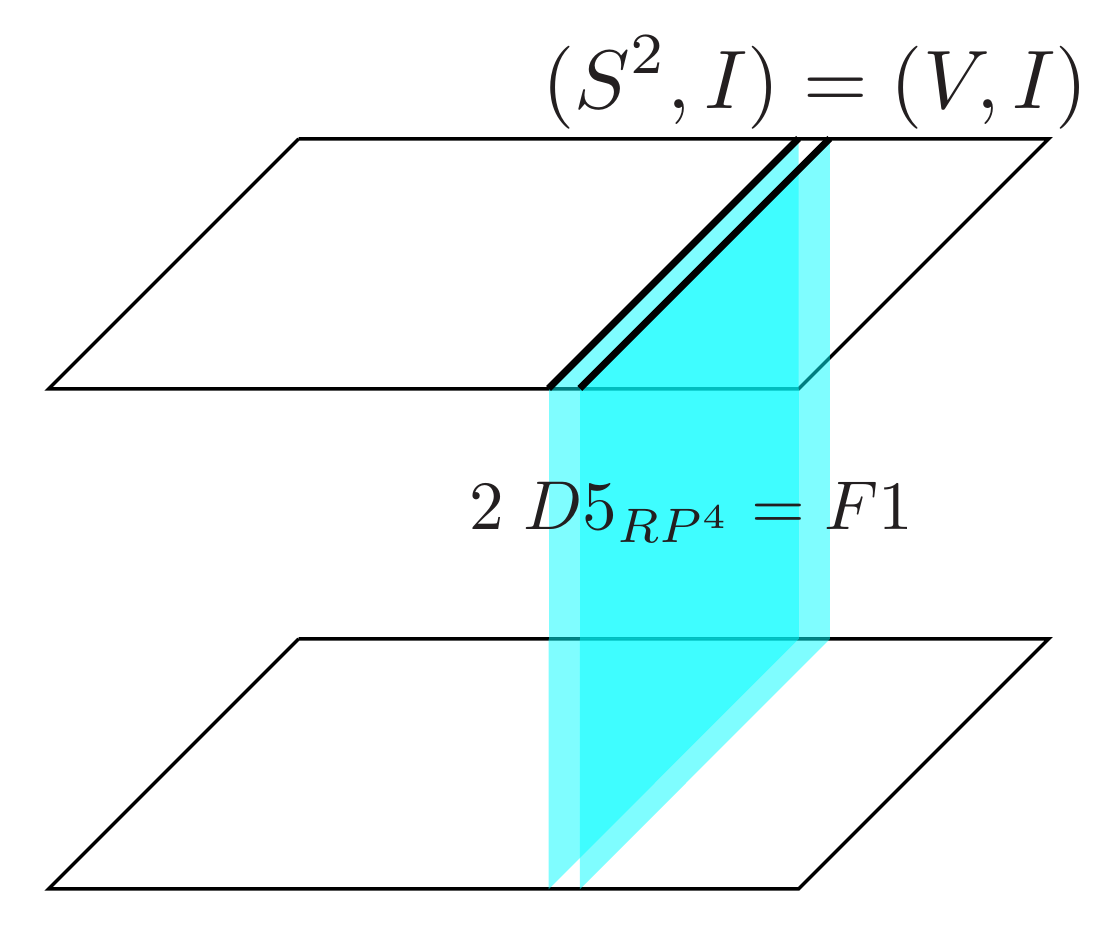}
\includegraphics[height=0.2\textwidth]{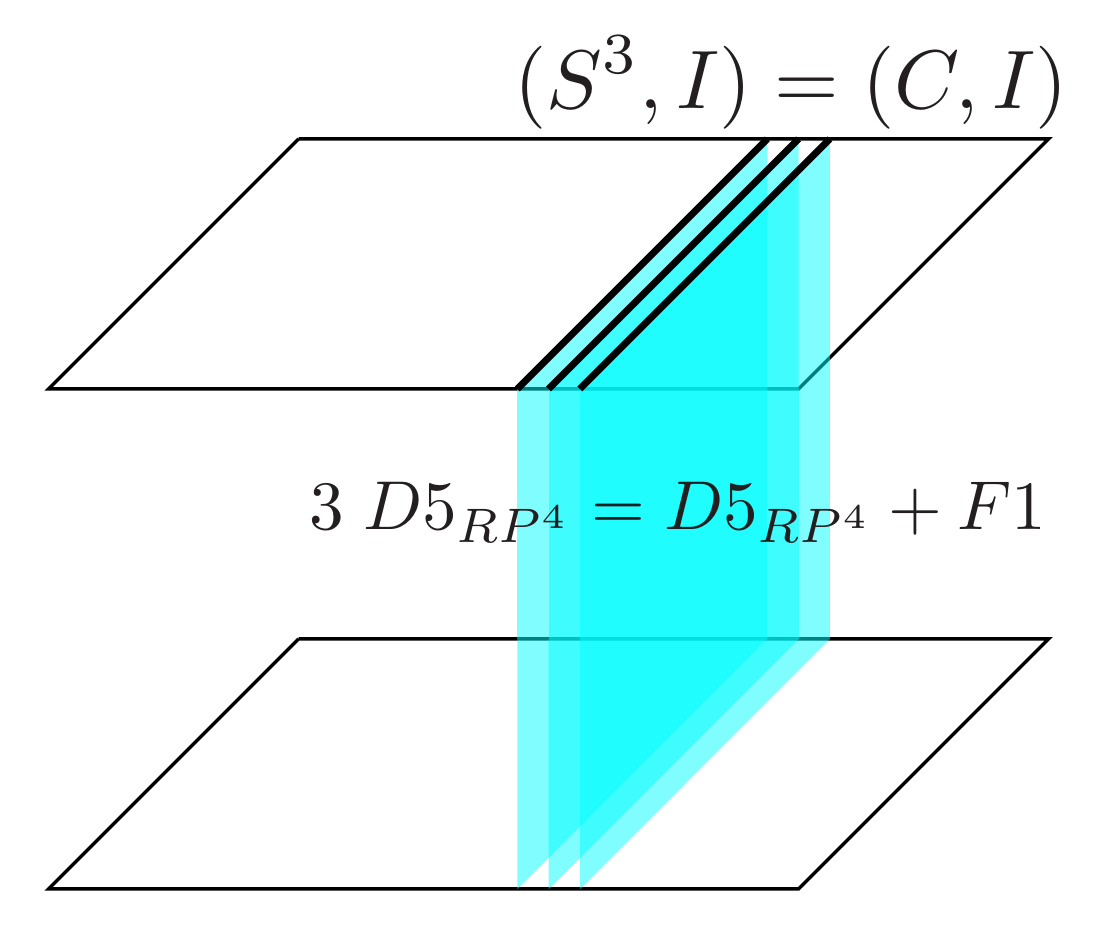}
\includegraphics[height=0.2\textwidth]{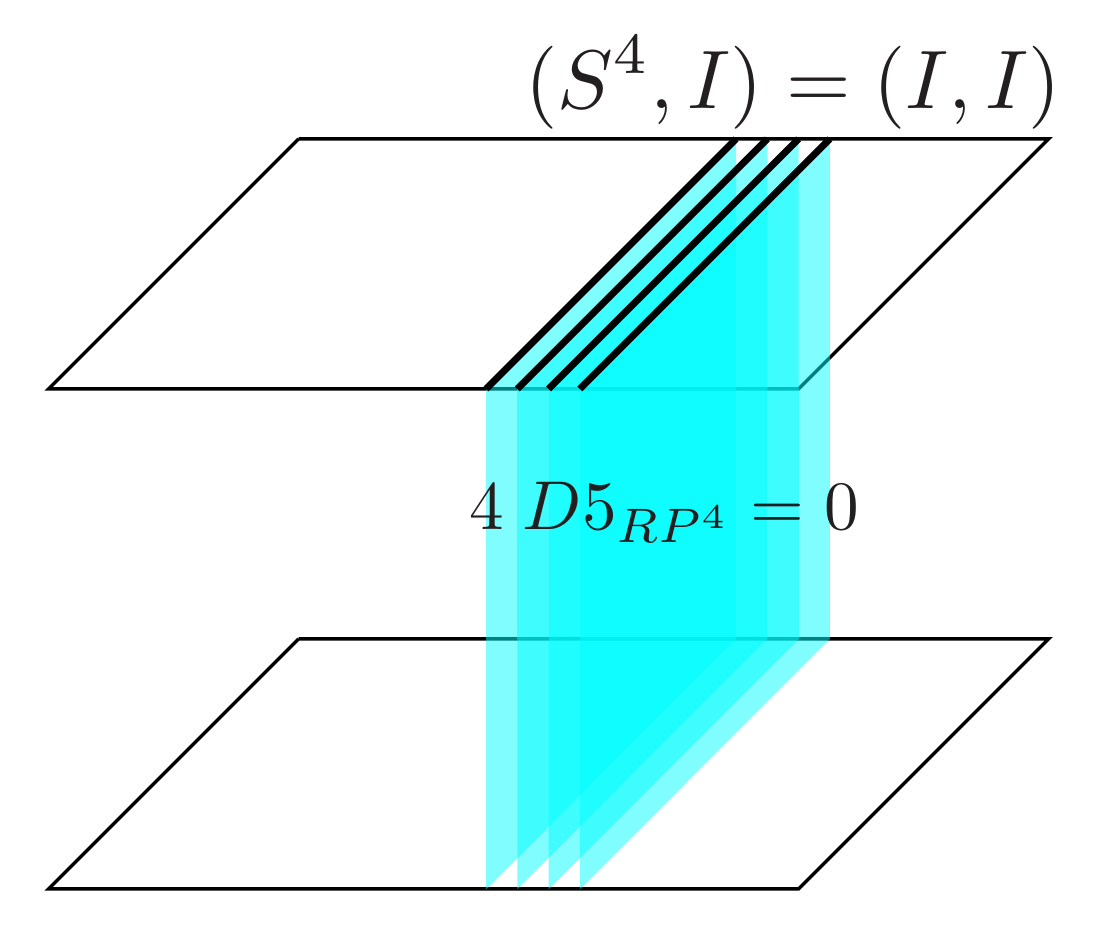}
\caption{Bulk description of the line operator spectrum of the $Spin(4k+2)$ theory.}
\label{SpinLines1}
\end{figure}

\noindent{\underline{$Spin(4k+2)$:}} 
The dual boundary condition fixes $\tilde{c}$, and therefore also $b = 2\tilde{c}$.
This means that only wrapped D5-branes and fundamental strings can end on the boundary of $AdS_5$,
and two wrapped D5-branes are equivalent to one fundamental string.
One wrapped D5-brane corresponds to an electric line in the spinor representation $S$, two wrapped D5-branes (or one string) 
to an electric line in the vector representation $V$, three wrapped D5-branes to an electric line in the spinor representation $C$,
and four correspond to a trivial line operator (see Fig.~\ref{SpinLines1}).
This is precisely the spectrum shown at the top of table~\ref{Table1}.

\medskip

\begin{figure}[h!]
\center
\includegraphics[height=0.2\textwidth]{Sline1}
\includegraphics[height=0.2\textwidth]{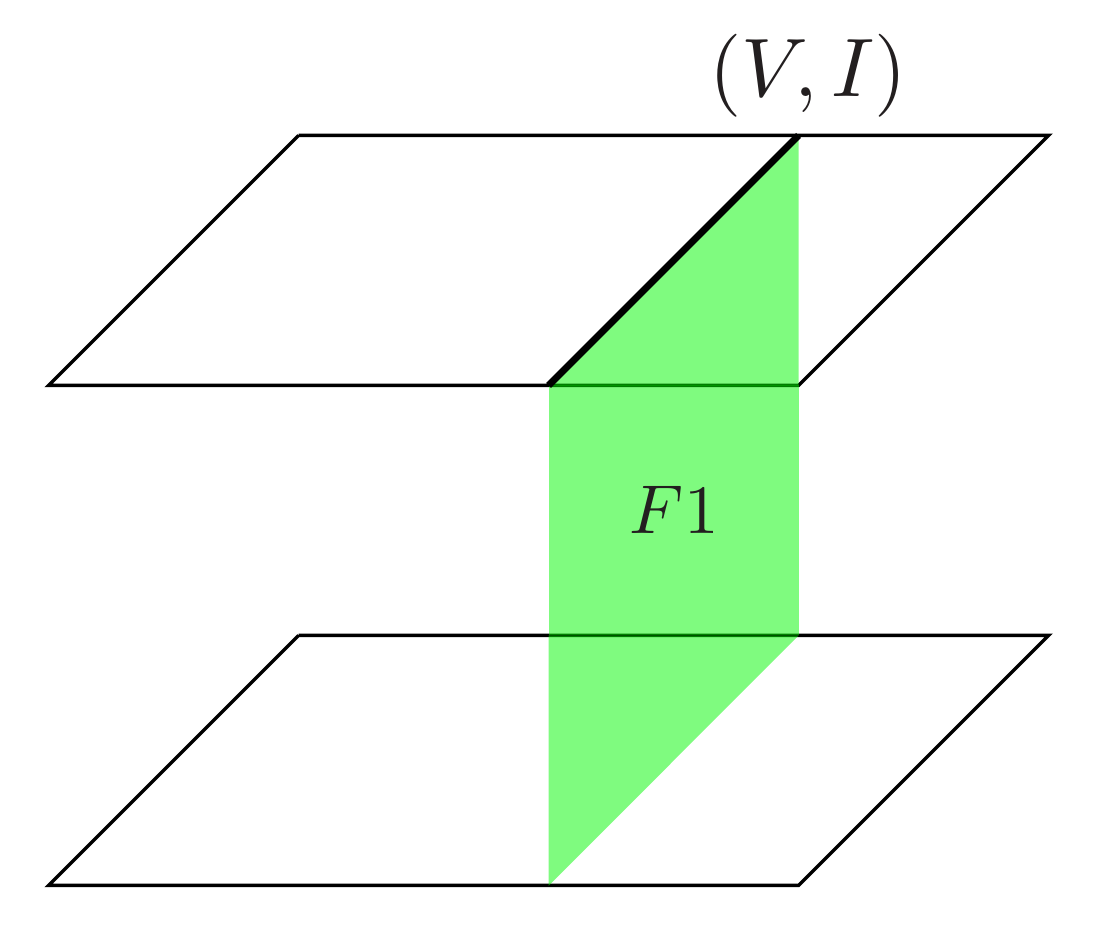}
\includegraphics[height=0.2\textwidth]{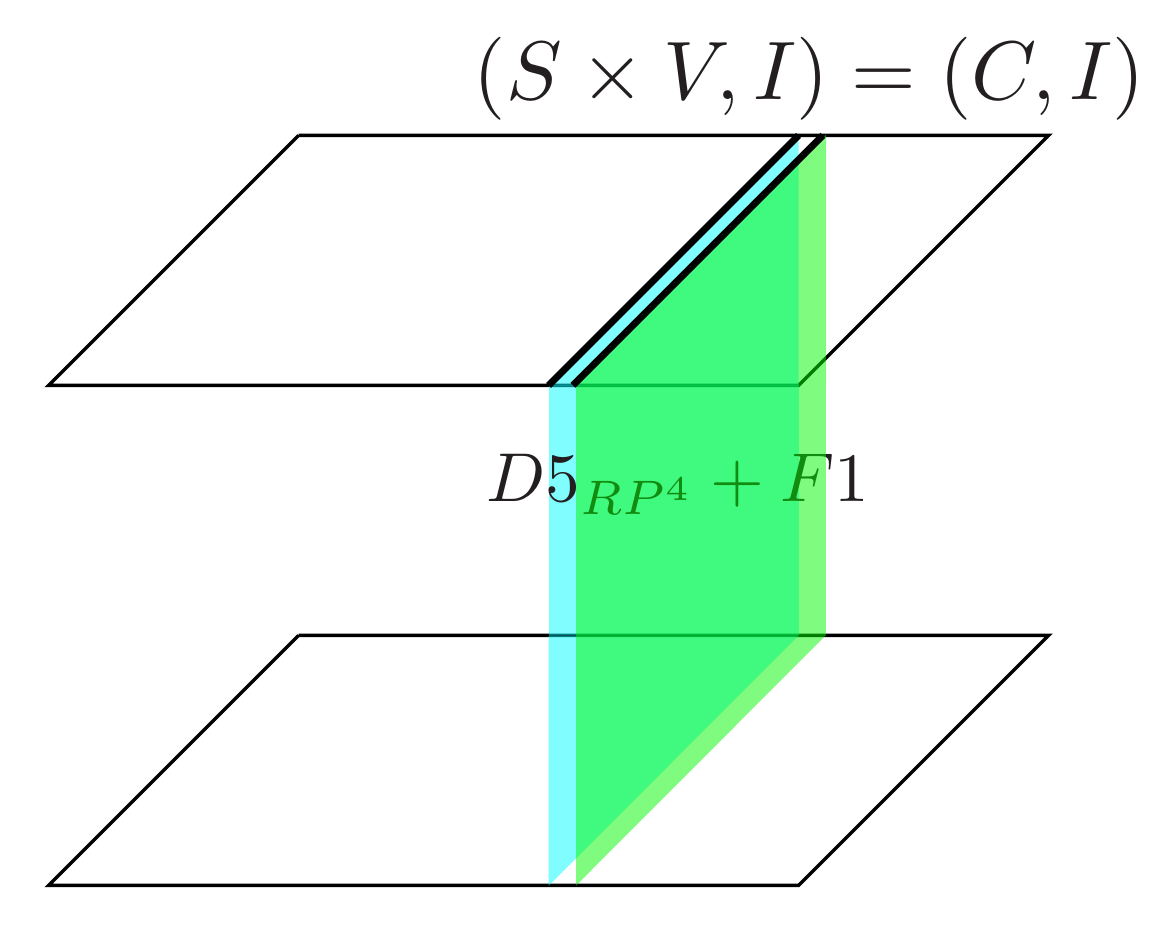}
\caption{Bulk description of the line operator spectrum of the $Spin(4k)$ theory.}
\label{SpinLines2}
\end{figure}

\noindent{\underline{$Spin(4k)$:}} 
For even $n$ the dual boundary conditions fix $\tilde{c}$ and $b$ separately. 
So as in our first example, both D5-branes and fundamental strings can end on the boundary, but here there is no relation between them.
In this case one wrapped D5-brane corresponds to electric $S$-line and two are trivial,
and one fundamental string corresponds to an electric $V$-line and two are trivial.
We can also consider a D5-F1 combination, which corresponds to an electric $C$-line (see Fig.~\ref{SpinLines2}).
This is precisely the spectrum shown at the top of table~\ref{Table2}.

\medskip

\noindent{\underline{$SO(2n)_0$:}}  The dual boundary conditions fix $b$ and $c$. For $n$ odd these are equivalent to $2\tilde{c}$ and $2\tilde{b}$.
This means that only the fundamental string (or equivalently pairs of D5-branes for odd $n$) and the D-string (or equivalently pairs of NS5-branes for odd $n$) 
can end on the boundary.
The spectrum of line operators therefore consists of electric and magnetic vectors, in agreement with the second line in both tables~\ref{Table1}, \ref{Table2}.

Note that screening by another brane does not play a role here. There is no ``baryon vertex" in $AdS_5\times \mathbb{R}P^5$.

\section{The $so(2n+1)$ and $sp(n)$ theories}

\subsection{One form symmetries and dualities}

In either case the center is $\mathbb{Z}_2$ and so the line operator charges $(z_e,z_m)$ take values in $\mathbb{Z}_2$.
For $so(2n+1)$ the electric charge is carried by the spinor representation $S$ and the magnetic charge by the vector representation $V$
of the GNO-dual $sp(n)$ algebra,
and vice versa for the algebra $sp(n)$.
We note also that $S\times S = I$ and $V \times V = I$.
Electric lines in the vector representation of $so(2n+1)$ are screened by the Pfaffian operator, which carries one loose gauge index in this case.
The Dirac pairing condition is given by 
\be
z_e z_m' - z_m z_e' = 0 \; \mbox{mod} \; 2 \,.
\ee
There are three different theories in either case, see Table~\ref{Table5}.
\begin{table}[h!]
\begin{center}
\begin{tabular}{|l|l|l|}
  \hline 
  theory & $(z_e,z_m)$  & $G^{(1)}$ \\
 \hline
  $Spin(2n+1)$ & $(S,I)^n$  & $\mathbb{Z}_2$ \\
   \hline
  $SO(2n+1)_0$ & $(I,V)^n$ & $\mathbb{Z}_2$ \\
  \hline
  $SO(2n+1)_1$ & $(S,V)^n$ & $\mathbb{Z}_2$ \\
   \hline\hline
  $Sp(n)$ & $(V,I)^n$ & $\mathbb{Z}_2$ \\
  \hline
  $(Sp(n)/\mathbb{Z}_2)_0$ & $(I,S)^n$ & $\mathbb{Z}_2$ \\
  \hline
  $(Sp(n)/\mathbb{Z}_2)_1$ & $(V,S)^n$ & $\mathbb{Z}_2$\\
  \hline
    \end{tabular}
 \end{center}
\caption{The $so(2n+1)$ and $sp(n)$ theories.}
\label{Table5}
\end{table}

\begin{figure}[h!]
\center
\includegraphics[width=1.0\textwidth]{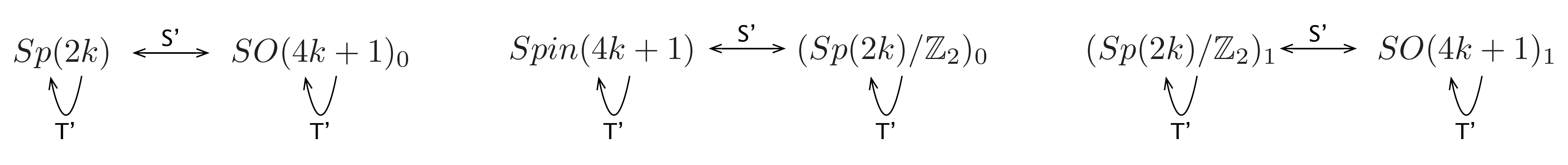} 
\caption{The duality orbits for $so(4k+1)$ and $sp(2k)$, reproduced from \cite{Aharony:2013hda}.}
\label{so(4k+1)Duality}
\end{figure}

\begin{figure}[h!]
\center
\includegraphics[width=1.0\textwidth]{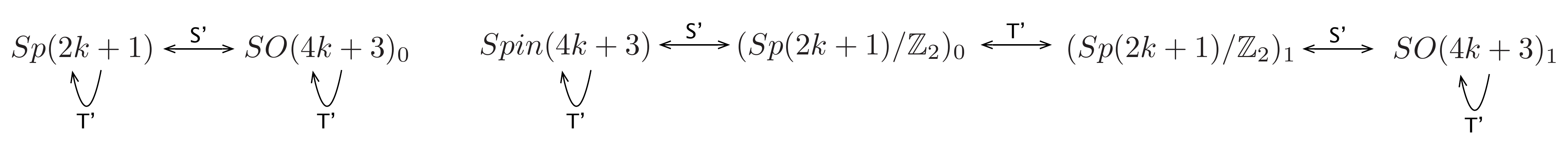} 
\caption{The duality orbits for $so(4k+3)$ and $sp(2k+1)$, reproduced from \cite{Aharony:2013hda}.}
\label{so(4k+3)Duality}
\end{figure}

The duality group in this case is not exactly $SL(2,\mathbb{Z})$, but rather the subgroup $[\Gamma_0(2) \rtimes \mathbb{Z}_4]/\mathbb{Z}_2 \subset SL(2,\mathbb{R})$
generated by \cite{Dorey:1996hx,Girardello:1995gf}
\be
\label{T'S'}
T' = \left(
\begin{array}{cc}
1 & 1 \\
0 & 1 
\end{array}
\right)
\quad
\mbox{and} 
\quad
S' = \left(
\begin{array}{cc}
0 & 1/\sqrt{2} \\
-\sqrt{2} & 0
\end{array}
\right) \,.
\ee
This acts in the standard way on the properly normalized complexified coupling,
\be 
\tau' \rightarrow \frac{a\tau' + b}{c\tau'+d} \,,
\ee
where
\be
\tau' 
= \frac{1}{2}\left(\frac{\theta}{2\pi} + \frac{4\pi i}{g^2}\right) \,.
\ee
In particular 
\be 
T': \; \theta \mapsto \theta + 4\pi \qquad  \mbox{and} \qquad {S'}: \; \tau' \mapsto \mbox{} - \frac{1}{2\tau'} \,.
\ee
This subtlety will be important when we discuss the string theory dual.
The duality orbits are slightly different for even and odd $n$, and are show in Figs.~\ref{so(4k+1)Duality},\ref{so(4k+3)Duality}.

\subsection{Holography}

The background $AdS_5\times \mathbb{R}P^5$ admits torsion-valued fluxes of the RR and NSNS 2-forms
given by $(\theta_{RR},\theta_{NS}) \in 
H^2(\mathbb{R}P^5,\tilde{\mathbb{Z}}) = \mathbb{Z}_2$ \cite{Witten:1998xy}.
The four possibilities correspond to the four O3-plane variants: $\mbox{O3}^{\pm}$ and $\widetilde{\mbox{O3}}^{\pm}$ \cite{Hanany:2000fq}.
The trivial flux background with $(\theta_{RR},\theta_{NS}) = (0,0)$, corresponding to the $\mbox{O3}^-$ plane, 
was discussed in the previous section. This is dual to the $so(2n)$ theories. 
The three other backgrounds with $(\theta_{RR},\theta_{NS})=(1/2,0)$, $(0,1/2)$, and $(1/2,1/2)$,
correspond, respectively to the $\widetilde{\mbox{O3}}^-$, $\mbox{O3}^+$, and $\widetilde{\mbox{O3}}^+$ planes,
and are dual, respectively, to the $so(2n+1)$, $sp(n)$, and $\widetilde{sp}(n)$ theories.
The $\widetilde{sp}(n)$ theories are based on the same algebra as the $sp(n)$ theories, 
but differ from them in the spectrum of magnetically charged states on the Coulomb branch \cite{Hanany:2000fq}.
In what follows we will concentrate on the $(1/2,0)$ and $(0,1/2)$ backgrounds, and briefly comment on the $(1/2,1/2)$
background and the dual $\widetilde{sp}(n)$ theories in a separate section.

While the $(0,0)$ background is invariant under the $SL(2,\mathbb{Z})$ symmetry of Type IIB string theory,
the other backgrounds transform as follows:
\be
\widetilde{\mbox{O3}}^- \stackrel{T}{\longleftrightarrow}  \widetilde{\mbox{O3}}^- \; , \; 
{\mbox{O3}}^+  \stackrel{T}{\longleftrightarrow}  \widetilde{\mbox{O3}}^+ \nonumber \\
\widetilde{\mbox{O3}}^-  \stackrel{S}{\longleftrightarrow}  {\mbox{O3}}^+ \; , \;
\widetilde{\mbox{O3}}^+  \stackrel{S}{\longleftrightarrow}  \widetilde{\mbox{O3}}^+ 
\ee
These transformations are clearly related to the duality transformations of the gauge theory,
however this relation is not as straightforward as in the previous cases.
We start by noting that the Type IIB axio-dilaton is related to the complexified coupling of the gauge theory as follows:
\be
\tau \equiv C_0 +ie^{-\Phi} = \frac{\theta}{2\pi} + \frac{4\pi i}{g^2} = 2{\tau'} \,.
\ee
The Type IIB supergravity action is given in the Einstein frame by
\be
S_{\rm IIB}={1\over 2\kappa^2_{10}}\int d^{10}x\sqrt{-G_E}\left[R_E-{\partial_{\mu}\bar{\tau}\partial^{\mu}\tau\over2({\rm Im}\tau)^2}
-{1\over 2}F_{\mu\nu\lambda}^i {\cal M}_i^j F_j^{\mu\nu\lambda}+\cdots\right]
\ee
where
\be
{\cal M}_i^j={1\over{\rm Im}\tau}\left(
\begin{array}{cc}
|\tau|^2 & -{\rm Re}\tau \\
-{\rm Re}\tau & 1
\end{array}
\right)\; , \;
F_3^i=(dB_2,dC_2) \,.
\ee
Consider the simple field redefinition
\be
(B_2',C_2') = \left(\sqrt{2} B_2, \frac{1}{\sqrt{2}} C_2\right) \,.
\ee
It follows that 
\be
S_{\rm IIB}[\tau,B_2,C_2] = S_{\rm IIB}[\tau',B_2',C_2'] \,.
\ee
Under the action $S'$
\be 
\tau' \rightarrow  -\frac{1}{2\tau'} = -\frac{1}{\tau} \quad \mbox{and} \quad
(B_2',C_2')  \rightarrow   \left(-\sqrt{2}C_2', \frac{1}{\sqrt{2}}B_2'\right) = (-C_2,B_2)\,,
\ee
and therefore 
\be
S_{\rm IIB}[\tau',B_2',C_2'] \rightarrow S_{\rm IIB}[-\frac{1}{\tau},-C_2,B_2]\,.
\ee
This is precisely what one would get by acting with the $S$ generator of $SL(2,\mathbb{Z})$, so we find that $S'=S$.
On the other hand under the action of $T'$ 
\be
\tau' & \rightarrow \tau' + 1 \quad \mbox{and} \quad 
(B_2',C_2') \rightarrow (B_2',C_2' + B_2') \,,
\ee
which translates to the action on $\tau$ and $(B_2,C_2)$
\be
\tau & \rightarrow \tau + 2 \quad \mbox{and} \quad 
(B_2,C_2) \rightarrow (B_2,C_2+ 2B_2) \,,
\ee
and therefore $T' = T^2$.
This implies in particular that the set of theories based on $sp(n)$ and the set based on $\widetilde{sp}(n)$,
dual to the $(0,1/2)$ and $(1/2,1/2)$ backgrounds, are separately invariant under $T'$.
As we mentioned above, we will discuss the $(1/2,1/2)$ background and the $\widetilde{sp}(n)$ theories in
a separate section.

The 5d CS action is the same as before (\ref{CSaction2}), but there are additional constraints on the fields due to the torsion NSNS or RR flux.
Recall that in the $(0,0)$ background we had the constraints (\ref{HolonomyRelations}) for odd $n$ originating from a condition on wrapped 5-branes. These constraints continue to hold in the other backgrounds.
The additional constraints
arise from additional conditions on wrapped branes \cite{Witten:1998xy}.
The first is a topological restriction on wrapping 5-branes: a D5-brane (NS5-brane) can wrap 
$\mathbb{R}P^4 \subset \mathbb{R}P^5$ only if $\theta_{NS}=0$ ($\theta_{RR}=0$).
More generally there is an obstruction to wrapping an odd number of D5-branes (or NS5-branes) if $\theta_{NS}\neq 0$ (or $\theta_{RR}\neq 0$).
We interpret this as a constraint on $\tilde{c}$ (or $\tilde{b}$) allowing only classes that are even multiples of the generator.
The second condition is on the D3-brane wrapping $\mathbb{R}P^3 \subset \mathbb{R}P^5$.
The RR (NSNS) flux gives rise to a tadpole which must be cancelled by attaching a fundamental string (D-string) to the D3-brane.
Equivalently, this means that a fundamental string can be screened by a wrapped D3-brane in the $(1/2,0)$ background,
and a D-string can be screened by a wrapped D3-brane in the $(0,1/2)$ background.
This implies an additional constraint $b=0$ in the $(1/2,0)$ background, 
and $c=0$ in the $(0,1/2)$ background.\footnote{These constraints suggest that the holonomies are K-theory classes 
rather than cohomology classes. Indeed in lifting cohomology to K-theory it is common that some classes may be obstructed while others may be trivialized. See for example \cite{Bergman:2001rp}.}

In either case we therefore have a single pair of canonically conjugate boundary holonomy variables valued in $\mathbb{Z}_2$.
In the $(1/2,0)$ background dual to the $so(2n+1)$ theories these are $c$ and $\tilde{c}$, corresponding to a magnetic vector and an electric spinor, respectively, with
\be
[c,\tilde{c}] = \pi i \;\; \mbox{mod} \;\; 2\pi i \,.
\ee
There are three allowed boundary conditions: $\tilde{c}=0$ corresponding to $Spin(2n+1)$, $c=0$ corresponding to $SO(2n+1)_0$,
and $c+\tilde{c}=0$ corresponding to $SO(2n+1)_1$.
For $sp(n)$ we only have $b$ and $\tilde{b}$, corresponding to an electric vector and a magnetic spinor, respectively, with 
\be
[b,\tilde{b}] = \pi i \;\; \mbox{mod} \;\; 2\pi i \,.
\ee
There are three allowed boundary conditions: $b=0$ corresponding to $Sp(n)$, $\tilde{b}=0$ corresponding to $(Sp(n)/\mathbb{Z}_2)_0$,
and $b+\tilde{b}=0$ corresponding to $(Sp(n)/\mathbb{Z}_2)_1$.
We summarize this in Table~\ref{Table7}.
\begin{table}[h!]
\begin{center}
\begin{tabular}{|l|l|}
  \hline 
  theory & boundary conditions \\
 \hline
  $Spin(2n+1)$ & $\tilde{c}=0$  \\
   \hline
  $SO(2n+1)_0$ & $c=0$  \\
  \hline
  $SO(2n+1)_1$ & $c + \tilde{c}=0$  \\
   \hline\hline
  $Sp(n)$ & $b=0$ \\
  \hline
  $(Sp(n)/\mathbb{Z}_2)_0$ & $\tilde{b}=0$ \\
  \hline
  $(Sp(n)/\mathbb{Z}_2)_1$ & $b + \tilde{b}=0$\\
  \hline
    \end{tabular}
 \end{center}
\caption{The boundary conditions dual to the $so(2n+1)$ and $sp(n)$ theories. The boundary conditions corresponding to the $\widetilde{sp}(n)$ theories
are identical to those of the corresponding $sp(n)$ theories (and to those of the $so(2n+1)$ theories).}
\label{Table7}
\end{table}

Next we consider the action of the field theory duality symmetry.
As we showed above the generator $S'$ is equal to the Type IIB $S$ transformation, $S'=S$.
This exchanges $\theta_{RR}$ and $\theta_{NS}$, and so exchanges the algebras $so(2n+1)$ and $sp(n)$.
The action on the holonomy variables is given by
\be
S: (b,c) \mapsto (c,-b) \quad \mbox{and} \quad (\tilde{b},\tilde{c}) \mapsto (-\tilde{c},\tilde{b})
\ee
This reproduces all the $S'$ maps in Figs.~\ref{so(4k+1)Duality} and \ref{so(4k+3)Duality}.
The second generator of the field theory duality group is given by $T'=T^2$. 
The fluxes $\theta_{NS}$ and $\theta_{RR}$ are invariant, so each algebra maps to itself, and the action on the holonomy variables is given by
\be
T^2: (b,c) \mapsto (b,c+2b) \quad \mbox{and} \quad (\tilde{b},\tilde{c}) \mapsto (\tilde{b} + 2\tilde{c},\tilde{c}) \,.
\ee
This reproduces all the $T'$ maps in Figs.~\ref{so(4k+1)Duality} and \ref{so(4k+3)Duality}.
Note in particular the different action of $T'$ on $(Sp(n)/\mathbb{Z}_2)_{0,1}$ for even $n$ and odd $n$.
This is due to the relations in (\ref{HolonomyRelations}).
The boundary condition dual to $(Sp(n)/\mathbb{Z}_2)_0$ is $\tilde{b}=0$.
Under $T'$ this becomes $\tilde{b}+2\tilde{c}=0$. Using (\ref{HolonomyRelations}) this becomes $\tilde{b}=0$ if $n$ is even,
and $\tilde{b}+b=0$ if $n$ is odd. In other words $(Sp(n)/\mathbb{Z}_2)_0$ maps to itself under $T'$ if $n$ is even (and similarly for
$(Sp(n)/\mathbb{Z}_2)_1$), and to $(Sp(n)/\mathbb{Z}_2)_1$ if $n$ is odd.

\subsection{Branes and line operators}

The discussion here can be relatively brief, since the spectrum of strings here is the same as in the previous section modulo
the additional constraints that we discussed above.
Fundamental strings and D-strings correspond to electric and magnetic vectors, respectively,
and wrapped D5-branes and NS5-branes correspond to electric and magnetic spinors, respectively.
In the background dual to the $so(2n+1)$ theories the NS5-brane is obstructed and the fundamental string is screened 
by a wrapped D3-brane. That leaves only the electric spinor and the magnetic vector as potential line operators.
The boundary condition dual to the $Spin(2n+1)$ theory allows only the wrapped D5-brane to end on the boundary, giving the electric spinor line.
The boundary condition dual to the $SO(2n+1)_0$ theory allows only the D-string to end on the boundary, giving the magnetic vector line.
The boundary condition dual to the $SO(2n+1)_1$ theory allows only the combination of a D-string and a wrapped D5-brane to end on the boundary, 
giving the dyonic $(S,V)$ line.
This is all in agreement with the spectrum of line operators in Table~\ref{Table5}.
A similar conclusion holds for the line operators of the $sp(n)$ theories, by exchanging the roles of the NS5-brane and D5-brane,
and of the fundamental string and the D-string.

\subsection{The $(1/2,1/2)$ background and the $\widetilde{sp}(n)$ theories}

Let us briefly comment on the $\widetilde{sp}(n)$ theories and their dual background with $(\theta_{RR},\theta_{NS})=(1/2,1/2)$.
From the gauge theory point of view the $sp(n)$ and $\widetilde{sp}(n)$ theories are related by ``half" of a $T'$ transformation, namely 
by a shift of $\theta \rightarrow \theta + 2\pi$, which is half its periodicity.
They should not therefore be regarded as different theories.
The holographic duals are related by a Type IIB $T$ transformation that takes $(\theta_{RR},\theta_{NS})=(0,1/2)$
to $(\theta_{RR},\theta_{NS})=(1/2,1/2)$.
The analysis of the allowed boundary conditions is similar to the $so(2n+1)$ and $sp(n)$ cases above.
In the case with both fluxes turned on the constraint from wrapped 5-branes requires the combination $\tilde{b} + \tilde{c}$
to be even, and the constraint from the wrapped 3-brane sets $b+c=0$.
%
%
Therefore we again have only one canonically conjugate pair, which we can take as either $(b,\tilde{b})$ or $(c,\tilde{c})$.
The identification of the boundary conditions dual to the different theories is the same as for $sp(n)$.


\section{Conclusions}

We have shown how the different consistent boundary conditions for 2-form gauge fields in Type IIB string theory on $AdS_5$
distinguish the different ${\cal N}=4$ Super-Yang-Mills theories based on the gauge algebras $su(N)$, $so(2n)$, $so(2n+1)$, and $sp(n)$,
in terms of their spectra of line operators and global one-form symmetries. We have also shown how the intricate structure of duality orbits
arises in the holographic description.
So far we have only discussed connected groups such as $SU(N)$, $SO(N)$, and $Spin(N)$.

An obvious natural extension would be to disconnected groups like $O(N)$, $Pin(N)$, and $\widetilde{SU}(N)$.\footnote{The group $\widetilde{SU}(N)$
is the principle extension of $SU(N)$ by the $\mathbb{Z}_2$ outer automorphism that exchanges the fundamental and antifundamental representations
\cite{Bourget:2018ond}.}
In the first two cases the relevant term in the 5d action is given by
\be
S_{CS}[A_1,A_3] = \frac{1}{\pi} \int_{AdS_5} A_1 \wedge dA_3 \,,
\ee
where $A_1$ and $A_3$ are given by the reductions of the RR 4-form $C_4$ on an $\mathbb{R}P^3$ and $\mathbb{R}P^1$ subspace of $\mathbb{R}P^5$,
respectively. For the theories discussed in this paper, namely $SO(N)$, $Spin(N)$, etc., $A_1$ is fixed at the boundary, and $A_3$ is free.
The one-form gauge field $A_1$ is dual to a $\mathbb{Z}_2$ 0-form symmetry that all these theories have.
In particular, for $N$ even it is the outer automorphism that exchanges the two Weyl spinors $S$ and $C$.
For $N=4k+2$ this is also identified with charge conjugation.
For $N$ odd it is just charge conjugation.
The theories with disconnected groups, namely $O(N)$, $Pin(N)$, etc., correspond to different boundary conditions, for example
fixing $A_3$ and allowing $A_1$ to be free.
The details of this should be worked out.
The case of $\widetilde{SU}(N)$ is also an interesting problem.

Another natural extension would be to non-Lagrangian $S$-fold theories such as the ${\cal N}=3$ theories of \cite{Garcia-Etxebarria:2015wns}.
The holographic duals of some of these theories are known \cite{Garcia-Etxebarria:2015wns,Aharony:2016kai}, and it would be interesting to understand the different possibilities corresponding
to the different choices of allowed boundary conditions.

In the $so(N)$ and $sp(N)$ cases we saw that various obstructions and relations were imposed on the discrete holonomy variables by brane dynamics.
These are familiar symptoms of the fact that in string theory fluxes are often classified in K theory rather than cohomology.
It would be interesting to make this more explicit.

\section*{Acknowledgments}

We thank O. Aharony, P-S Hsin, S.~Razamat, T.~Sakai and G.~Zafrir for useful discussions. The work of Oren Bergman is supported in part by the Israel Science Foundation under grant No. 1390/17. The work of Shinji Hirano is supported in part by the
National Natural Science Foundation of China under Grant No.12147219.

\appendix

\section{An explicit isomorphism between (\ref{OneFormSymmetry1}) and (\ref{OneFormSymmetry2})}
\label{Generators}

Let $x,y$ denote the generators of $\mathbb{Z}_{k'}$ and $\mathbb{Z}_{N/gcd(k',\ell)}$, respectively,  in (\ref{OneFormSymmetry1}),
and $u,v$ denote the generators of $\mathbb{Z}_{N/gcd(k,k',\ell)}$ and $\mathbb{Z}_{gcd(k,k',\ell)}$,
respectively, in (\ref{OneFormSymmetry1}).
The quotient in (\ref{OneFormSymmetry1}) imposes the relation $y^k=x^\ell$.
In general we have
\be
u = x^p y^q \; , \; v = x^s y^t \,,
\ee
where $p,q,s,t\in\mathbb{Z}$.
For $v$ we require that $v^{gcd(k,k',\ell)} = 1$. The general solution is
\be 
\label{vAnsatz}
s= \frac{\beta k' - \alpha \ell}{\mbox{gcd}(k,k',\ell)} \; , \; t = \frac{\alpha k}{\mbox{gcd}(k,k',\ell)} \,.
\ee
where $\alpha,\beta \in\mathbb{Z}$.
The group generated by $u$ is in general $\mathbb{Z}_{N/gcd(k',pk + q\ell)}$:
\be
u^{N/gcd(k',pk+q\ell)} =  (x^py^q)^{N/gcd(k',pk+q\ell)}=x^{k'(pk+q\ell)/gcd(k',pk+q\ell)}=1 \,.
\ee
We therefore require the pair $(p,q)$ to satisfy
\be
\label{pqCondition}
\mbox{gcd}(k',pk+q\ell) = \mbox{gcd}(k,k',\ell) \,.
\ee
The proof that such a pair of integers exists will be given below (in fact one can take $q=1$),
but let us now proceed assuming that it does.
In order for the map to be an isomorphism we require that it be invertible, namely that we can express $x$ and $y$ in terms
of $u$ and $v$,
\be 
x = u^a v^b \; , \; y = u^c v^d
\ee
where $a,b,c,d \in \mathbb{Z}$.
If we assume that the four integers $p,q,s,t$ satisfy
\be
\label{pqstCondition}
qs - pt = 1
\ee
it is easy to see that the following choice for $a,b,c,d$ does the job
\be
(a,b,c,d) = (-t,q,s,-p) \,.
\ee
Is our assumption in (\ref{pqstCondition}) valid? Substituting in the expressions for $s$ and $t$ from (\ref{vAnsatz}) and using (\ref{pqCondition}) gives
\be
qs - pt = \beta \left[\frac{k'}{\mbox{gcd}(k',pk+q\ell)}\right] - \alpha \left[\frac{pk+q\ell}{\mbox{gcd}(k',pk+q\ell)}\right]
\ee
The two integers in square brackets are clearly co-prime, and therefore we are guaranteed by Bezout's identity
that there exists a coprime pair $(\alpha,\beta)$ such that this equal to 1, so (\ref{pqstCondition}) is indeed satisfied.

The $SL(2,\mathbb{Z})$ matrices of (\ref{SmithNormalForm}) are then given by
\begin{align}
U=\left(
\begin{array}{cc}
p & 1 \\
{-\beta k'+\alpha\ell\over{{\rm{gcd}}(k,k',\ell)}} & {-\alpha k\over{{\rm{gcd}}(k,k',\ell)}}
\end{array}
\right)\;\; , \;\;
V=\left(
\begin{array}{cc}
{pk+q\ell\over{{\rm{gcd}}(k,k',\ell)}} & {k'\over{{\rm{gcd}}(k,k',\ell)}} \\
-\beta & -\alpha
\end{array}
\right) \,.
\end{align}

\subsection{Proof of (\ref{pqCondition})}

All that remains to show is that there exists a pair of integers $(p,q)$ that satisfy (\ref{pqCondition}).
Let us simplify the notation a bit by defining
\be
(m,n) \equiv \left(\frac{\mbox{gcd}(k',\ell)}{\mbox{gcd}(k,k',\ell)},\frac{k}{\mbox{gcd}(k,k',\ell)}\right) \; , \; 
(g,h) \equiv \left(\frac{k'}{\mbox{gcd}(k',\ell)},\frac{\ell}{\mbox{gcd}(k',\ell)}\right) \,.
\ee
Clearly $\mbox{gcd}(m,n) = \mbox{gcd}(g,h) = 1$. We can rewrite the conjecture (\ref{pqCondition}) as
\be 
\mbox{gcd}(gm,pn + qhm) =1 \,.
\ee
We can rewrite the LHS as
\be 
\mbox{gcd}(gm,pn+qhm) &=& \mbox{gcd}(g,pn+qhm) \mbox{gcd}(m,pn+qhm) \nonumber \\[5pt]
&=& \mbox{gcd}(g,pn+qhm)  \mbox{gcd}(m,pn) \\[5pt]
&=&  \mbox{gcd}(g,pn+qhm)  \mbox{gcd}(m,p) \,. \nonumber
\ee
We therefore need to show that there exists a pair of integers $(p,q)$ such that
\be
\mbox{gcd}(m,p) = \mbox{gcd}(g,pn+qhm)  =1 \,.
\ee
First let us take $q=1$. Then we can show that there exists an integer $p$ satisfying these two conditions iteratively as follows.
If $\mbox{gcd}(g,m)=1$, and therefore $\mbox{gcd}(g,hm)=1$, we can just take $p=g$.
Otherwise, define $(g_1,m_1) = (g,m)/\mbox{gcd}(g,m)$, and then
\be
\mbox{gcd}(g,pn+hm) = 
\mbox{gcd}(g_1,pn + hm_1\mbox{gcd}(m,g)) \, \mbox{gcd}(\mbox{gcd}(m,g),p) \,.
\ee
Then if $\mbox{gcd}(g_1,\mbox{gcd}(g,m)) =1$, and therefore $\mbox{gcd}(g_1,hm_1 \mbox{gcd}(g,m)) =1$, we can take $p=g_1$.
Otherwise define $(g_2,m_2) = (g_1,\mbox{gcd}(g,m))/\mbox{gcd}(g_1,\mbox{gcd}(g,m))$, and then
\be
 \mbox{gcd}(g_1,pn + hm_1\mbox{gcd}(g,m))  &=& \mbox{gcd}(g_2,pn + hm_1m_2\mbox{gcd}(g_1,\mbox{gcd}(m,g))) \nonumber \\
 &&\mbox{} \times\mbox{gcd}(\mbox{gcd}(g_1,\mbox{gcd}(m,g)),p) \,.
 \ee
Then if $\mbox{gcd}(g_2,\mbox{gcd}(g_1,\mbox{gcd}(g,m))) = 1$, and therefore 
$\mbox{gcd}(g_2, hm_1m_2\mbox{gcd}(g_1,\mbox{gcd}(m,g)))  = 1$, we can take $p=g_2$.
We repeat this process until at the $N$'th step we get
\be
\mbox{gcd}(g_N,\mbox{gcd}(g_{N-1},\mbox{gcd}(g_{N-2},\mbox{gcd}(g_{N-3},\mbox{gcd}(g_{N-4},\cdots , \mbox{gcd}(g,m))\cdots ) = 1 \,,
\ee
at which point the solution is $p=g_N$.

\section{Counting the maximal charge lattices for $so(4k)$}

The mutual locality conditions \eqref{so(4k)Locality} can be expressed more succinctly as
\begin{align}\label{Z2inner}
\Tr\left[\hat{Z}^{T}i\sigma_2\hat{Z}'\right]=0\quad\mbox{mod}\quad 2
\end{align}
where $\hat{Z}$ and $\hat{Z}'$ are $2\times 2$ matrix-valued vectors of the form
\begin{align}
\hat{Z}:=
\begin{pmatrix}
z_1 & z_3\\
z_2 & z_4
\end{pmatrix}\ .
\end{align}
Thus the mutual locality conditions can be interpreted as orthogonality of two vectors which are null with respect to the inner product defined on the LHS of \eqref{Z2inner}. 
It is more illuminating to represent the vector $\hat{Z}$ in the following basis:
\begin{equation}
\begin{aligned}
\hat{Z}&=z_1{\mathbb{I}_2+\sigma_3\over 2}
+z_2{\sigma_1-i\sigma_2\over 2}
+z_3{\sigma_1+i\sigma_2\over 2}
+z_4{\mathbb{I}_2-\sigma_3\over 2}\\
&\equiv z_1\hat{i}_1+z_2\hat{i}_2
+z_3\hat{i}_3+z_4\hat{i}_4\ .
\end{aligned}
\end{equation}
Using $\sigma_i\sigma_j=i\epsilon_{ijk}\sigma_k$, we see that the two pairs of bases, $\{\hat{i}_1, \hat{i}_2\}$ and $\{\hat{i}_3, \hat{i}_4\}$, are not orthogonal to each other. 
This implies that there are only four mutually orthogonal pairs of bases: $\{\hat{i}_1, \hat{i}_3\}\equiv{\bf Pair}_{13}$, $\{\hat{i}_1, \hat{i}_4\}\equiv{\bf Pair}_{14}$, $\{\hat{i}_2, \hat{i}_3\}\equiv{\bf Pair}_{23}$, and $\{\hat{i}_2, \hat{i}_4\}\equiv {\bf Pair}_{24}$ in contrast to 4d vectors with the standard inner product for which there are $\mbox{}_4C_2=6$ mutually orthogonal pairs.  In particular, there are no triplet or quartet of mutually orthogonal bases. For example, the triplet $\{\hat{i}_1, \hat{i}_2, \hat{i}_3\}$ are not mutually orthogonal. This thus shows that the charge lattices are two-dimensional.

\medskip
How can we find all 15 pairs of mutually orthogonal pairs of vectors? 
The simplest are the 4 pairs of bases, ${\bf Pair}_{13}$, ${\bf Pair}_{14}$, ${\bf Pair}_{23}$, and ${\bf Pair}_{24}$.
The linear combinations of two of them generate 4 new independent pairs of (non-basis) vectors: ${\bf Pair}_{1(3+4)}$, ${\bf Pair}_{2(3+4)}$, ${\bf Pair}_{3(1+2)}$, and ${\bf Pair}_{4(1+2)}$.
Similarly, the linear combinations of these new pairs generate a new independent pair of vectors: ${\bf Pair}_{(1+2)(3+4)}$. 
Now, due to the mod 2 property of orthogonality, even though the pairs $\{\hat{i}_1, \hat{i}_2\}$ and $\{\hat{i}_3, \hat{i}_4\}$ are not individually orthogonal, the sums of them can be. This yields 2 more pairs, ${\bf Pair}_{(1+3)(2+4)}$ and ${\bf Pair}_{(1+4)(2+3)}$.
From these 2 pairs, we can generate the remaining 4 pairs: since $\hat{i}_2$ is orthogonal to itself and $\hat{i}_4$, we can construct a new pair ${\bf Pair}_{(1+2+3)(2+4)}$ from ${\bf Pair}_{(1+3)(2+4)}$. Note that ${\bf Pair}_{(1+2+3)(2+4)}\sim {\bf Pair}_{(1+2+3+(2+4))(2+4)}={\bf Pair}_{(1+3+4)(2+4)}$.
A similar consideration yields the remaining 3 pairs, ${\bf Pair}_{(1+3)(1+2+4)}$ from ${\bf Pair}_{(1+3)(2+4)}$
and ${\bf Pair}_{(1+4)(1+2+3)}$, ${\bf Pair}_{(1+2+4)(2+3)}$ from ${\bf Pair}_{(1+4)(2+3)}$.

\end{document}